\documentclass[11pt]{article}
\usepackage{pifont}
\usepackage{amsfonts}
\usepackage{subfigure}

\usepackage{amssymb}
\usepackage{amsmath}
\usepackage{epsfig}

\begin{document}
\newtheorem{theorem}{Theorem}
\newtheorem{corollary}{Corollary}
\newtheorem{conjecture}{Conjecture}
\newtheorem{definition}{Definition}
\newtheorem{lemma}{Lemma}
\newtheorem{algorithm}{Algorithm}
\newtheorem{remark}{Remark}
\newtheorem{idea}{Idea}
\newtheorem{observation}{Observation}

\newcommand{\define}{\stackrel{\triangle}{=}}

\pagestyle{empty}

\def\sDoF{\overline{\mbox{\normalfont \scriptsize DoF}}}

\def\QED{\mbox{\rule[0pt]{1.5ex}{1.5ex}}}
\def\proof{\noindent{\it Proof: }}

\date{}

\title{Beyond One-Way Communication: Degrees of Freedom of Multi-Way Relay MIMO Interference Networks}
\author{\normalsize  Chenwei Wang\\
        {\small DOCOMO Innovations Inc., Palo Alto, CA 94304}\\
      {\small \it E-mail~:~cwang@docomoinnovations.com}\\
       }
\maketitle

\thispagestyle{empty}
\begin{abstract}

We characterize the degrees of freedom (DoF) of multi-way relay MIMO interference networks. In particular, we consider a wireless network consisting of 4 user nodes, each with $M$ antennas, and one $N$-antenna relay node. In this network, each user node sends one independent message to each of the other user nodes, and there are no direct links between any two user nodes, i.e., all communication must pass through the relay node. For this network, we show that the symmetric DoF value per message is given by $\max(\min(\frac{M}{3},\frac{N}{7}),\min(\frac{2M}{7},\frac{N}{6}))$ normalized by space dimensions, i.e., piecewise linear depending on $M$ and $N$ alternatively. While the information theoretic DoF upper bound is established for every $M$ and $N$, the achievability relying on linear signal subspace alignment is established in the spatially-normalized sense in general. In addition, by deactivating 4 messages to form a two-way relay MIMO $X$ channel, we also present the DoF result in the similar piecewise linear type. The central new insight to emerge from this work is the notion of inter-user signal subspace alignment incorporating the idea of network coding, which is the key to achieve the optimal DoF for multi-way relay interference networks. Moreover, this work also settles the feasibility of linear interference alignment that extends the feasibility framework from one-way to multi-way relay interference networks.
\end{abstract}

\allowdisplaybreaks

\newpage

\section{Introduction}

In wireless networks with multiple user nodes, concurrent transmissions give rise to competition for channel resources between multiple information flows. How to deal with interference caused by concurrent transmissions is they key to understand the fundamental capacity limit of wireless networks. Among a variety of interference management schemes, recently, a new multiplexing technique called {\em interference alignment} was proposed to study the degrees of freedom (DoF) of communication networks \cite{Jafar_FTn}. While many interference alignment schemes have been designed for a wide variety of multiuser networks, the central common insight is to align interference as much as possible, while the desired signal of each receiver can still be distinguishable. So far, the DoF characterizations available are almost for one-way communication networks, i.e., each node in the network either sends or demands messages, but not both. However, in general communication networks, most user nodes are likely to both send and demand messages. In contrast to one-way networks, such a class of communication networks are referred to as multi-way networks. For multi-way communication networks, channel capacity, or even the DoF characterizations, remain widely open.

\subsection{The Problem}

For multi-way communication networks, recently Lee \emph{et. al.} studied the DoF of a 3-user MIMO $Y$ channel in \cite{Lee_Lim_MIMOY} where each user equipped with 2 antennas sends one independent message to each of the other user nodes via the help of a 3-antenna relay only, and there are no direct links between any two user nodes. While their achievable scheme can achieve 1 DoF per message, i.e., 2 DoF per user, which is the single-user DoF upper bound, what is interesting is the new idea of \emph{alignment for network coding}. Specifically, they first designed linear beamforming vectors at each user, so that every two pairwise symbols, i.e., the symbols of two pairwise users, each sending to and demanding from the other, are aligned along the same vector at the relay. After receiving the signal, the relay is able to resolve the 3 linear combinations, each of which is the sum of two pairwise symbols. Afterwards, they applied a reciprocal approach to design beamforming vectors at the relay and each user, so that each user finally only sees two linear combinations, each of which is the sum of one of its desired symbols and one of its own transmitted symbols. Since each user's own symbols are available at itself as side information, it can subtract the signal carrying its own symbols to obtain an equivalent clean single-user channel, and then to decode its desired symbols. Thus, the phrase of alignment for network coding essentially means signal alignment at the relay and network coding at each user for decoding.

Regarding the idea of alignment for network coding in \cite{Lee_Lim_MIMOY}, we are interested in the question whether we can directly apply it to the setting with more than 3 users. To simply the problem as much as possible, we consider a symmetric but general setting where there are 4 $M$-antenna users, each sending one independent message to each of the others via an $N$-antenna relay only, and there are no direct links between any two users, as shown in Fig. \ref{fig:system}, where $M$ and $N$ can take arbitrary positive integers. We
will study the symmetric DoF per message of this network. Actually, this problem is challenging for two reasons. First, since $M,N$ can take arbitrary values, every two users may have not project a common intersection at the vector space at the relay. In the signal alignment terminology, it implies that \emph{one-to-one alignment} is impossible, and we have to use \emph{one-to-many alignment}, by meaning that one symbol lies in the subspace \emph{jointly} spanned by many other symbols. However, it is quite challenging to design an
efficient achievability scheme. Second, as we will explain later, for every user, the number of interfering messages increase from 2 in the 3-user setting to 6 in this work. Thus, identifying the complex interplay between the subspaces at the relay carrying each message is quite interesting and nontrivial.

\subsection{Prior Work}

Of the vast amount of literature on beyond one-way relay interference networks, the most closely related to this work are references \cite{Lee_Lim_MIMOY, Lee_KMIMOY, Lee_MIMO_2way, Yang_2way, Wang_Jafar_2way, Sezgin_Y, Yuan_xjtu}. In particular, following the idea of alignment for network coding in \cite{Lee_Lim_MIMOY}, the DoF of several multi-way networks were studied in \cite{Lee_KMIMOY, Lee_MIMO_2way, Yang_2way}. However, they all only tackled with networks with special number of antennas, so that the signals carrying every two pairwise symbols can be aligned along the same vector at the relay. Thus, their results, requiring one-to-one alignment only, are established in a relatively straightforward manner. In the absence of one-to-one alignment, Wang \emph{et. al.} studied the DoF of the 2-pair and 3-pair two-way relay MIMO interference channel \cite{Wang_Jafar_2way}, based on the idea of inter-pair signal subspace alignment. Compared to \cite{Wang_Jafar_2way}, this work also needs one-to-many alignment. However, since the total number of messages increase from 6 in \cite{Wang_Jafar_2way} to 12 in this work, how to identify the interplay between the subspaces projected from each user at the relay, and thus the DoF characterizations become much more challenging. In \cite{Sezgin_Y}, Chaaban \emph{et. al.} characterized the DoF of a general 3-user relay MIMO $Y$ channel, on the top of the particular model in \cite{Lee_Lim_MIMOY}, where each node is equipped with arbitrary number of antennas. However, for this channel, again, one-to-one alignment is sufficient to achieve the DoF upper bound. Regarding the problem we study in this paper, recently, Yuan \emph{et. al.} showed an achievable DoF result in \cite{Yuan_xjtu}, essentially based on the idea of inter-pair signal subspace alignment in \cite{Wang_Jafar_2way}. However, their result is the DoF achievability only, and it is not clear whether their result is tight, thus leaving this problem still open in general.

\subsection{Contribution}

In this paper, we show that the symmetric DoF value per message is piecewise linear depending on $M$ and $N$ alternatively. To establish this result, we provide both the information theoretic DoF converse and the DoF achievability. While our DoF converse is established for every $M$ and $N$, the DoF achievability relying on linear signal vector alignment is established in the spatially-normalized sense in general. We remind the reader that the similar observations were also illustrated in \cite{Wang_Gou_Jafar_3userMxN} for studying the DoF of the 3-user MIMO interference channel. However, as we will explain in detail later, the ideas behind both the DoF converse and the DoF achievability appear to be different. As a byproduct of this work, we also study the DoF of a two-way relay MIMO $X$ channels, and we present the DoF result in the similar piecewise linear type. The key to establish the new results of this work is \emph{how to design an efficient scheme using the idea of inter-user signal subspace alignment for network coding}. Compared to the recent work \cite{Yuan_xjtu} where only the achieved DoF are presented, we provide both the information theoretic DoF upper bound and the DoF inner bound. In particular, we show that the result in \cite{Yuan_xjtu} is not tight when $3/8\leq M/N \leq 1/2$, by identifying a gap between their achieved DoF and the upper bound developed in this work, and further closing the gap with a new achievable scheme. For example, consider the network $(M,N)=(3,7)$, which is also the most interesting case for this work. It was shown in \cite{Yuan_xjtu} that each message can achieve $7/8$ DoF. However, we show in this paper that each message can achieve 1 DoF, which are also the information theoretic DoF upper bound.

\subsection{Significance}

We believe that our contribution is interesting for three reasons.

First, as mentioned earlier in this section, the open problem studied in this paper has attracted much attention these years, and finally we settle this problem in this work.

Second, as we explain in this paper, for both the DoF converse and the DoF achievability, we essentially translate the original network to a one-way one-hop channel, from the users to the relay, by imposing additional constraints at the relay. Contrary to conventional one-hop one-hop channels where every message is either undesired or desired but not both at a receiver, in this work, since all communication must pass through the relay, every message is both desired (to its desired decoder) and undesired (to other decoders) at the relay. Thus, the imposed additional constraints at the relay, is the key to formulate the multi-way relay network to a one-way one-hop channel, which provides a lens to study the DoF of multi-way relay networks.

Finally, as all existing work on the feasibility of linear interference alignment are for one-way one-hop wireless networks, in this paper we also settle the feasibility of linear interference alignment for two multi-way relay MIMO interference networks. To the best of
our knowledge, this is the first work establishing the feasibility of linear interference alignment if the network is beyond one-hop.

\section{System Model}\label{sec:system}

\begin{figure}[!t] \centering
\includegraphics[width=3.6in]{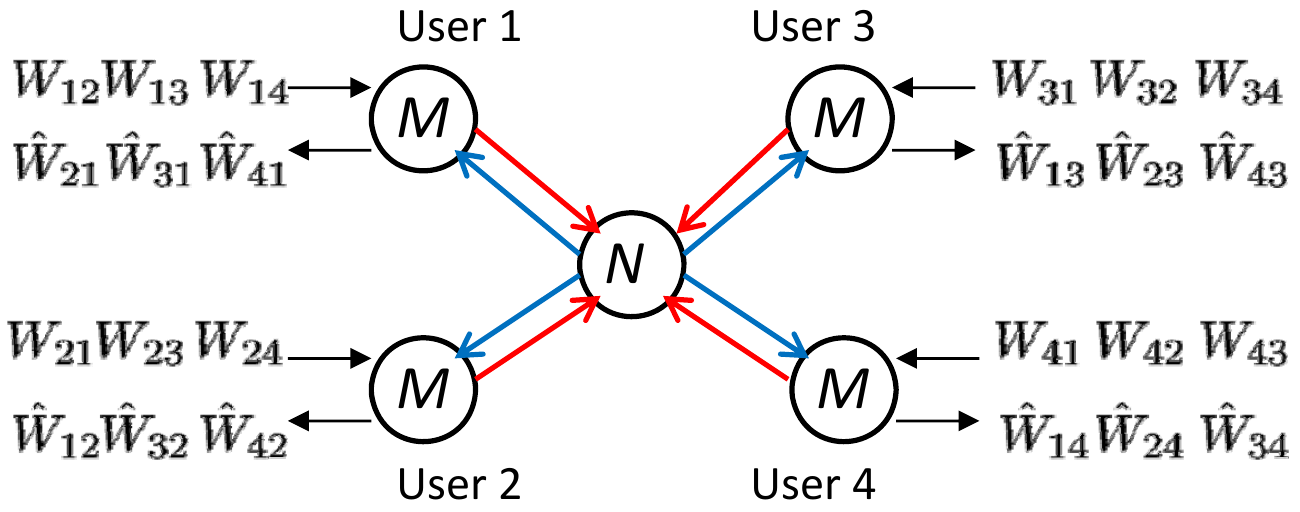}\vspace{-0.1in}
\caption{The System Model}\label{fig:system}
\end{figure}

Consider a wireless network where there are 4 user nodes, each with $M$ antennas, and one relay node with $N$ antennas. As shown in Fig. \ref{fig:system}, each user $k\in\{1,2,3,4\}\triangleq \mathcal{K}$ sends one independent message $W_{kj}$ to each of the other 3 users where $j\in\mathcal{K}\setminus\{k\}$ via the help of the relay node only, and there are no direct links between any two users. We denote by ${\bf H}_k$ the $N\times M$ channel matrix from user $k$ to the relay, and $\bar{{\bf H}}_k$ the $M\times N$ channel matrix from the relay to user $k$. Moreover, we assume that the channel coefficients are independently drawn from continuous distributions, and stay constant during the entire transmission once they are drawn. We also assume that global channel knowledge is available at every node in the network. Notice that our results are valid regardless of whether the channel matrices ${\bf H}_k$ and $\bar{{\bf H}}_k^T$ are identical or not for each user $k$. In this work, we assume that all the 5 nodes work in the full-duplex mode, i.e., they can hear and transmit simultaneously\footnote{If all nodes work in the half-duplex mode, then the DoF results we show in this paper will be scaled by a factor $1/2$ due to normalization to time.}.

Since the relay hears from all users, its received signal vector at time $t\in\mathbb{Z}^+$ is given by:
\begin{eqnarray}
Y_R(t)=\sum_{k\in\mathcal{K}} {\bf H}_k X_k(t)+Z_R(t)
\end{eqnarray}
where $X_k(t)$ is the complex-valued $M\times 1$ signal vector sent from user $k$, the $N\times 1$ column vector $Z_R(t)$ represents the i.i.d. circularly symmetric complex additive white Gaussian noise (AWGN) at the relay and $Z_R(t)\sim \mathcal{CN}(0,{\bf I}_N)$. At the user node, each user hears from the relay only. Thus, the received signal vector at user $k$ at time $t$ is given by:
\begin{eqnarray}
Y_k(t)= \bar{{\bf H}}_k X_R(t)+Z_k(t)
\end{eqnarray}
where $X_R(t)$ is the complex-valued $N\times 1$ signal vector sent from the relay, the $M\times 1$ column vector $Z_k(t)$ represents the AWGN and $Z_k(t)\sim \mathcal{CN}(0,{\bf I}_M)$. In addition, we assume that the transmitted signal vectors from all nodes $i\in\mathcal{K}\cup\{R\}$ satisfy the average power constraint $\frac{1}{T}\sum_{t=1}^T \mathbb{E}[\|X_i(t)\|^2]\leq P$ for $T$ channel users.

In this work, we will study the DoF of the following two settings:
\begin{itemize}
\item{} All the 12 messages are active, to form a 4-user relay MIMO $Y$ channel, which is a natural extension of the 3-user relay MIMO $Y$ channel. We refer to this case as the \emph{all unicast} setting.

\item{} Among the 12 messages, we set $W_{12}=W_{21}=W_{34}=W_{43}=\emptyset$, to form a two-way relay MIMO $X$ channel, because every user on both the left-hand-side and the right-hand-side sends one independent message to every user on the other side. We refer to it as the \emph{multiple unicast} setting.
\end{itemize}
We denote by $R=R_{kj}$ the symmetric rate of the message $W_{kj}$ and the corresponding symmetric DoF metric is denoted as $d$. The rate and the DoF definitions follow from their standard definitions in information theory. In addition, the definition of the spatially normalized DoF metric, to avoid special channel structures and to keep generic channels for the DoF achievability, is introduced in \cite{Wang_Gou_Jafar_3userMxN}, and we omit it here to avoid repetition. Basically, it means that we scale the number of antennas at each node by a factor $q$ so that the resulting DoF value $q\cdot d$ is an integer, much like $q$ symbol extensions over the time/frequnecy domains. Based on all available results so far, from the DoF perspective, DoF normalization to spatial extensions (without special channel structures), is similar to normalization to time/frequency (with block diagonal structures). That is, scaling time/frequency/spatial resources for a network also scales its DoF with the same factor.

{\it Notations:} We use $a$, $A$ and ${\bf A}$, ${\bf I}_m$ to denote a scalar, a column vector, a matrix and the $m\times m$ identity matrix, respectively. Also, ${\bf A}^T$, ${\bf A}^H$ stand for the transpose and the conjugate transpose of the matrix ${\bf A}$, respectively. In addition, we denote by $A(m:n)$ the sub-column vector whose entries are picked from the $m^{th}$ to the $n^{th}$ entries of the vector $A$ sequentially. Moreover, we use $\epsilon(x)$ to represent any function so that $\lim_{x\rightarrow \infty}\epsilon(x)/x=0$.

\section{Main Results}\label{sec:results}

We state our main DoF results and illustrate the main insights behind the results in this section.

\subsection{All Unicast: The 4-user MIMO $Y$ Channel}

{\bf Definition 1:} Define the following quantity $d_Y^*=\max(\min(\frac{M}{3},\frac{N}{7}),\min(\frac{2M}{7},\frac{N}{6}))$, or equivalently
\begin{eqnarray*}
d_Y^*=\left\{\begin{array}{lll}M/3,&&0<M/N\leq 3/7,\\N/7,&&3/7<M/N\leq 1/2,\\2M/7,&&1/2<M/N\leq 7/12,\\N/6,&&7/12<M/N.\end{array}\right.
\end{eqnarray*}

\begin{lemma} \label{lemma:y_ob}
(DoF Converse) For the 4-user relay MIMO $Y$ channel that we defined in Section \ref{sec:system}, the information theoretic DoF per message are upper bounded by $d\leq d_Y^*$.
\end{lemma}

{\it Proof:} The proof is presented in Section \ref{sec:y_ob}.\hfill\QED

\begin{lemma} \label{lemma:y_ib}
(DoF Achievability) For the 4-user relay MIMO $Y$ channel that we defined in Section \ref{sec:system}, each message can achieve $d_Y^*$ spatially normalized DoF almost surely.
\end{lemma}

{\it Proof:} The proof is presented in Section \ref{sec:y_ib}.\hfill\QED

\begin{theorem}\label{thm:y}
For the 4-user relay MIMO $Y$ channel defined in Section \ref{sec:system}, each message has $d_Y^*$ spatially normalized DoF.
\end{theorem}

{\it Proof:} The proof follows directly from Lemma \ref{lemma:y_ob} and Lemma \ref{lemma:y_ib}. \hfill\QED

\begin{figure}[!b] \centering \vspace{-0.1in}
\includegraphics[width=3.6in]{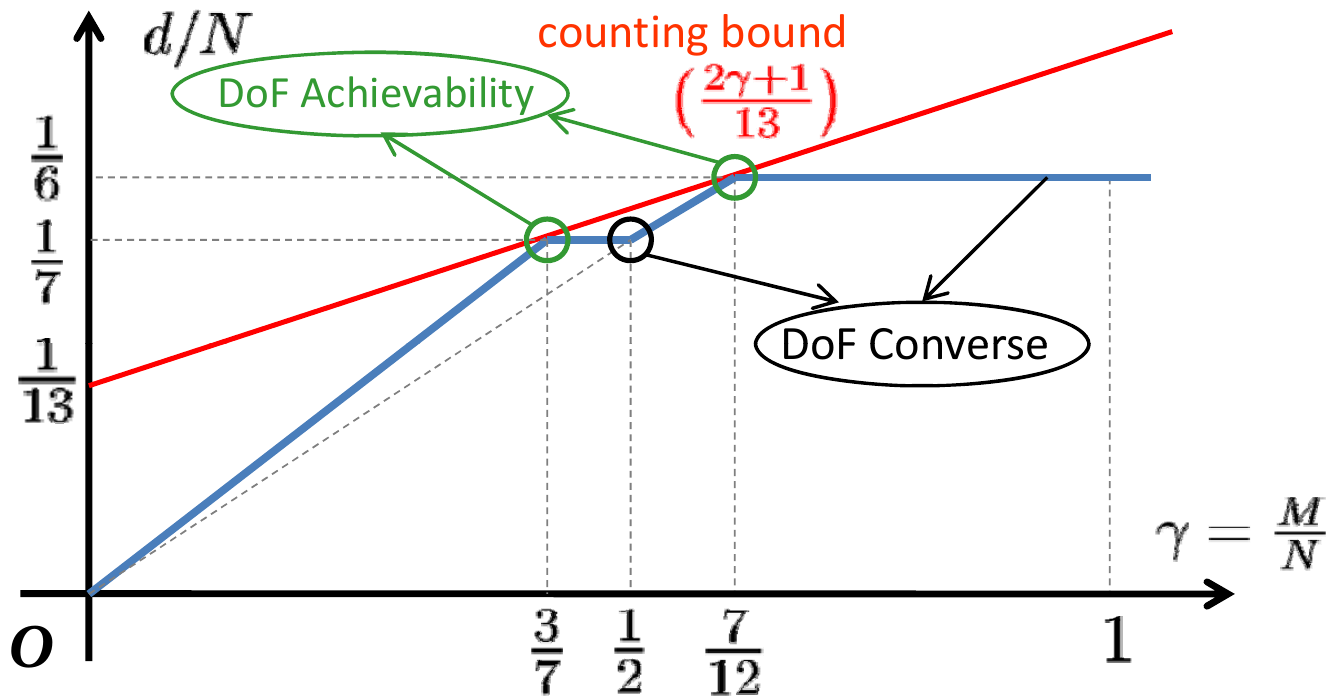}\vspace{-0.15in}
\caption{DoF per message for the 4-user relay MIMO $Y$ channel}\label{fig:resultY}
\end{figure}

We plot the DoF result implied by Theorem \ref{thm:y} in Fig. \ref{fig:resultY}. It can be seen that the DoF value per message is piecewise linear depending on $M$ and $N$ alternatively, which implies that there are antenna dimension redundancies at either each user or the relay node. Recall that the similar observation was recently also made for the 3-user $M\times N$ MIMO interference channel by Wang \emph{et.al.} in \cite{Wang_Gou_Jafar_3userMxN}. However, the tools used to obtain these results appear to be different. In \cite{Wang_Gou_Jafar_3userMxN}, the DoF achievability based on the new idea of subspace alignment chains, only needs one-to-one alignment, and the length of subspace alignment chains depends on the values of $M,N$ only. When $\gamma=M/N$ increases, we keep adding new symbols to subspace alignment chains one by one. In contrast, in this work, we have to use one-to-many alignment when $\gamma=M/N$ falls into some regimes, and thus there does not exist the concept of subspace alignment chains. In particular, when $\gamma=M/N$ increases, e.g., from $M/N=3/7$ to $M/N=7/12$, the type of the achievable scheme changes from one-to-many alignment to one-to-one alignment. While it is clear how to design an one-to-one alignment scheme, the most challenge of this work comes from how to design an efficient one-to-many alignment scheme which can achieve the DoF upper bound.

Similar to \cite{Wang_Gou_Jafar_3userMxN}, we conclude the antenna redundancies observations from Fig. \ref{fig:resultY}. Specifically, if and only if $M/N=1/2$, both $M$ and $N$ include redundant antenna dimensions. If and only if $M/N\in \{3/7,7/12\}$, neither $M$ nor $N$ contains any redundant dimensions. Thus, intuitively the DoF converse originates from $M/N=1/2$ (within the black circles) and $M/N\geq 7/12$, and the DoF achievability originates from $M/N=3/7,7/12$ (with green circles). Once we finish the proofs at these points, we can use the similar idea with additional efforts to solve every case between every two adjacent transition points.

\subsection{Multiple Unicast: The two-way Relay MIMO $X$ Channel}

{\bf Definition 2:} Define the following quantity $d_X^*=\max(\min(\frac{M}{2},\frac{N}{5}),\min(\frac{2M}{5},\frac{N}{4}))$, or equivalently
\begin{eqnarray*}
d_X^*=\left\{\begin{array}{lll}M/2,&&0<M/N\leq 2/5,\\N/5,&&2/5<M/N\leq 1/2,\\2M/5,&&1/2<M/N\leq 5/8,\\N/4,&&5/8<M/N.\end{array}\right.
\end{eqnarray*}

\begin{lemma} \label{lemma:x_ob}
(DoF Converse) For the two-way relay MIMO $X$ channel that we defined in Section \ref{sec:system}, the information theoretic DoF per message are upper bounded by $d\leq d_X^*$.
\end{lemma}

{\it Proof:} Since the proof is similar to that for Lemma \ref{lemma:y_ob}, we defer the proof into Appendix \ref{app:xob}.\hfill\QED

\begin{lemma} \label{lemma:x_ib}
(DoF Achievability) For the two-way relay MIMO $X$ channel that we defined in Section \ref{sec:system}, each message can achieve $d_X^*$ spatially normalized DoF almost surely.
\end{lemma}

{\it Proof:} Since the proof is similar to that for Lemma \ref{lemma:y_ib}, we defer the proof into Appendix \ref{app:xib}. \hfill\QED

{\it Remark:} We point out that the most interesting and nontrivial DoF achievability at $M/N=2/5$ is essentially much simpler than the achievability at $M/N=3/7$ of the all unicast setting.

\begin{theorem} \label{thm:x}
For the two-way MIMO $X$ channel defined in Section \ref{sec:system}, each message has $d_X^*$ spatially normalized DoF.
\end{theorem}

{\it Proof:} The proof follows directly from Lemma \ref{lemma:x_ob} and Lemma \ref{lemma:x_ib}. \hfill\QED

{\it Remark:} Recently, this problem was also studied by Xiang \emph{et. al.} in \cite{Tao_sjtu} where they only demonstrated that the total number of DoF is upper bounded by $2\min(2M,N)$, i.e., the DoF per message are upper bounded by $\frac{\min(2M,N)}{4}$, and this bound can be achieved if $M/N\geq 5/8$. Thus, among the 4 regimes implied by Theorem \ref{thm:x}, their DoF converse only covers the first and last regimes, and their DoF achievability is for the last regime only, thus leaving the DoF characterization of this network open in general. In this paper, our DoF converse is for every $M$ and $N$, and our DoF achievability is for every $M$ and $N$ as well but in the sense of spatial extensions.

\begin{figure}[!t] \centering 
\includegraphics[width=3.6in]{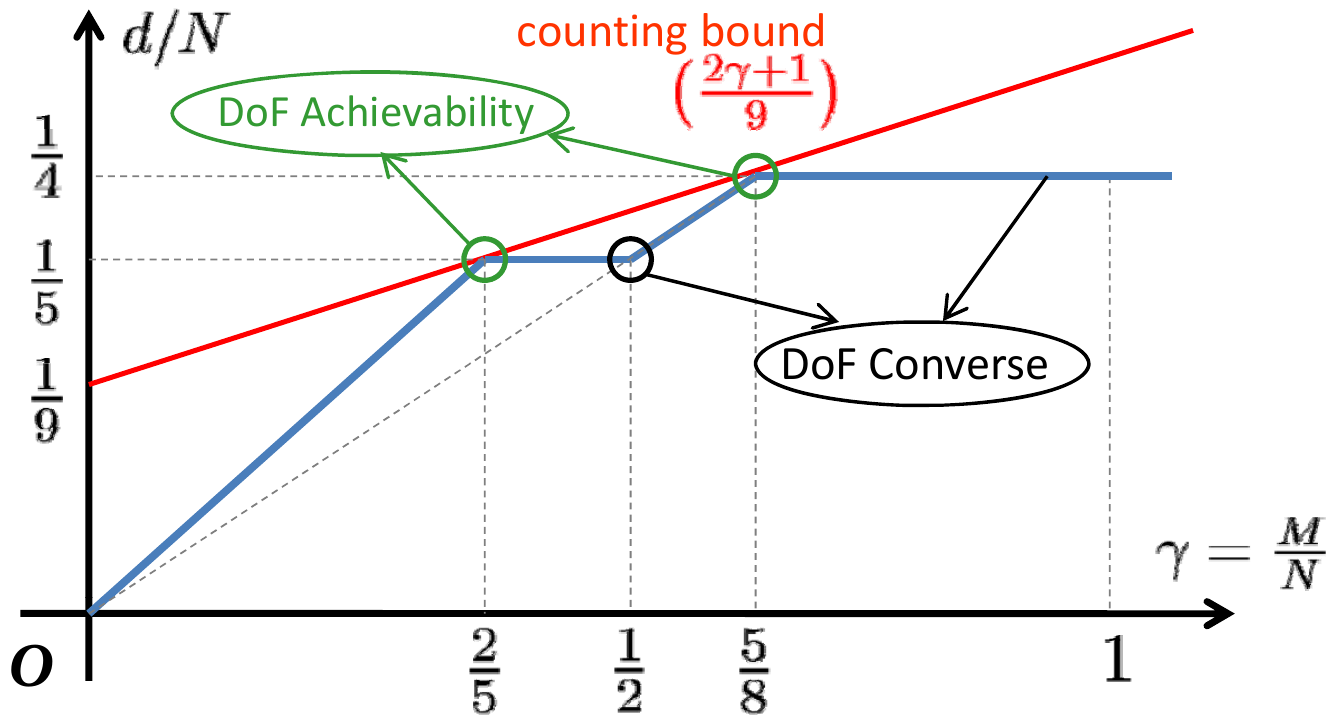}\vspace{-0.1in}
\caption{DoF per message for the two-way relay MIMO $X$ channel}\label{fig:resultX}
\end{figure}

We also plot the DoF result implied by Theorem \ref{thm:x} in Fig. \ref{fig:resultY}. Theorem \ref{thm:x} implies the similar observations such as piecewise linear, antenna dimension redundances.

\subsection{Feasibility of Linear Interference Alignment}

For the all unicast setting and the multiple unicast setting, we also plot the DoF counting bounds in Fig. \ref{fig:resultY} and Fig. \ref{fig:resultX}, respectively, represented by the red straight line in each figure. The DoF counting bound, first originated from the work by Cenk \emph{et. al.} in \cite{Cenk_Gou_Jafar} for studying the feasibility of linear interference alignment of MIMO interference channels, identifies the system into proper and improper. In this work, we follow from the similar counting approach to produce the DoF counting bounds $d\leq \frac{2M+N}{13}$ for the all unicast setting, and $d\leq \frac{2M+N}{9}$ for the multiple unicast setting. The development of the DoF counting bounds is deferred into Appendix \ref{app:county} and  Appendix \ref{app:countx}. Since the observations of the two settings are similar, let us consider the all unicast setting as an example. From Fig. \ref{fig:resultY}, it turns out \emph{for this network, improper is infeasible, many proper systems are infeasible, and if the information theoretic DoF upper bound is lower than the counting bound, then linear beamforming schemes are sufficient to achieve the information theoretic DoF upper bound}. These observations, already available for one-way one-hop MIMO interference channels in many prior works such as \cite{Wang_Gou_Jafar_3userMxN}, are verified to exist in the beyond one-hop network as well. Following from our similar observations that we have made for the one-way MIMO interference channel \cite{Wang_Gou_Jafar_3userMxN}, we also present the feasible DoF values, i.e., the DoF achieved by linear beamforming schemes without symbol extensions, for the two networks we study in this work.
\begin{theorem}\label{theorem:feasibility}
For the 4-user relay MIMO $Y$ channel defined in Section \ref{sec:system}, the DoF demand per message, $d$, are feasible with linear interference alignment if and only if $d\leq \lfloor d_Y^* \rfloor$ DoF.
\end{theorem}
{\it Proof:} The proof is deferred to Appendix \ref{app:feasibility}. \hfill\QED

\begin{theorem}
For the two-way relay MIMO $X$ channel defined in Section \ref{sec:system}, the DoF demand per message, $d$, are feasible with linear interference alignment if and only if $d\leq \lfloor d_X^* \rfloor$ DoF.
\end{theorem}
{\it Proof:} Since this channel model is a special setting of the 4-user relay MIMO $Y$ channel, the proof for this theorem essentially follows from the proof of Theorem \ref{theorem:feasibility}, and thus we omit it in this paper. \hfill\QED

\section{DoF Converse: Proof of Lemma 1}\label{sec:y_ob}

As implied by Lemma 1, there are 4 linear pieces depending on $M$ and $N$ alternatively. Let us consider the regime $M/N\leq 3/7$ first. Since each user sends 3 messages and each user is equipped with $M$ antennas only, the total number of DoF per user cannot be larger the than single-user DoF bound $M$. Thus, the DoF per message are upper bounded by $M/3$.

Next, we consider the remaining regimes sequentially. Note that each user only hears from the relay, thus the received signal at each user is a degraded version of what the relay sends. Since each user is able to decode its 3 desired messages with its 3 own messages as side information, if a genie provides that side information to the relay, then the relay is able to decode the messages desired at that user as well. By doing so, the DoF converse for the one-hop Multiple Access Channel (MAC) is also the DoF converse the original two-hop channel. Thus, we will use the genie-aided approach for the one-hop MAC to establish the DoF converse.

\subsection{$M/N\geq 7/12 \Longrightarrow d\leq N/6$}

In order for the reader to understand the DoF converse proof easily, we will first build the proof by using the linear dimension counting approach, and then translate it into the information theoretic proof.

Suppose that a genie provides to the relay the messages $\mathcal{G}=\{W_{12}, W_{13}, W_{14}, W_{32}, W_{42}, W_{34}\}$. Now, let us count the total number of dimensions contributed by the other 6 messages $W_{21}$, $W_{31}$, $W_{41}$, $W_{23}$, $W_{24}$, $W_{43}$. First, since user 1 can decode $W_{21},W_{31},W_{41}$ with its own 3 messages $W_{12},W_{13},W_{14}$ as side information, which are also provided to the relay by the genie, the relay is able to decode $W_{21},W_{31},W_{41}$ as well. Thus, the 3 messages $W_{21},W_{31},W_{41}$ contribute a total of $3d$ dimensions, which have to be linearly independent with the interfering dimensions contributed by the rest 3 messages $W_{23},W_{24},W_{43}$. Next, note that $W_{23},W_{43}$ can be both decoded at user 3 where $W_{31}, W_{32}, W_{34}$ are available as side information. Since $W_{32}, W_{34}$ are already provided by the genie at the relay, and $W_{31}$ can also be decoded first and thus again available at the relay, the relay is able to decode the messages $W_{23},W_{43}$ as well. Hence, $W_{23},W_{43}$ contribute other $2d$ dimensions. Finally, consider the message $W_{24}$ desired at user 4 where $W_{41}, W_{42}, W_{43}$ are available as side information. Again, $W_{42}$ is already available at the relay by the genie, and $W_{41},W_{43}$ can be first decoded and thus available at the relay as well. Therefore, $W_{42}$ contributes the other $d$ dimensions. So far, all the 6 messages $W_{21}$, $W_{31}$, $W_{41}$, $W_{23}$, $W_{24}$, $W_{43}$ contribute a total of $3d+2d+d=6d$ dimensions which cannot be larger than the number of antennas at the relay node. Thus, we have $6d\leq N$ to produce the desired upper bound $d\leq N/6$.

{\it Remark:} Among the 6 messages $W_{21}$, $W_{31}$, $W_{41}$, $W_{23}$, $W_{24}$, $W_{43}$, it can be seen that every message is not a paired message of any other. That is, $W_{ij}$ and $W_{ji}$ do not both appear among these 6 messages. Intuitively, to guarantee every message to be decoded at its own desired receiver, we need to protect $d$ dimensions for each pairwise messages $(W_{ij}, W_{ji})$. Thus, all those 6 messages occupy a total of $6d$ dimensions, so as to produce the desired DoF upper bound.

In the following, we translate the linear dimension counting approach into the information theoretic statement. Let a genie provide to the relay the messages $\mathcal{G}=\{W_{12}, W_{13}, W_{14}, W_{32}, W_{42}, W_{34}\}$. Then we consider the sum rate of the 3 messages desired at user 1:
\begin{eqnarray}
n(R_{21}+R_{31}+R_{41})
&\!\!\!\!\leq \!\!\!\!&  I(W_{21},W_{31},W_{41};Y_1^n|W_{12},W_{13},W_{14})+\epsilon(n)\label{eqn:Yob1_fano}\\
&\!\!\!\!\leq \!\!\!\!&  I(W_{21},W_{31},W_{41};Y_R^n|W_{12},W_{13},W_{14})+\epsilon(n)\label{eqn:Yob1_dp}\\
&\!\!\!\!\leq \!\!\!\!&  I(W_{21},W_{31},W_{41};Y_R^n,\mathcal{G}|W_{12},W_{13},W_{14})+\epsilon(n)\label{eqn:Yob1_genie}\\
&\!\!\!\!= \!\!\!\!&  I(W_{21},W_{31},W_{41};\mathcal{G}|W_{12},W_{13},W_{14})+I(W_{21},W_{31},W_{41};Y_R^n|\mathcal{G})+\epsilon(n)\label{eqn:Yob1_chain}\\
&\!\!\!\!= \!\!\!\!& I(W_{21},W_{31},W_{41};Y_R^n|\mathcal{G})+\epsilon(n)\label{eqn:Yob1_s}
\end{eqnarray}
where (\ref{eqn:Yob1_fano}) follows from the Fano's inequality; (\ref{eqn:Yob1_dp}) is obtained via the data processing inequality because $Y_R-X_R-Y_1$ forms a Markov chain; (\ref{eqn:Yob1_genie}) is obtained because adding genie signals does not reduce the capacity region; and (\ref{eqn:Yob1_s}) follows from the fact that the first term in (\ref{eqn:Yob1_chain}) is zero.

Also, consider the sum rate of the 2 messages $W_{23},W_{43}$ desired at user 3:
\begin{eqnarray}
n(R_{23}+R_{43})&\!\!\!\!\leq \!\!\!\!&  I(W_{23},W_{43};Y_3^n|W_{31},W_{32},W_{34})+\epsilon(n)\\
&\!\!\!\!\leq \!\!\!\!&  I(W_{23},W_{43};Y_R^n|W_{31},W_{32},W_{34})+\epsilon(n)\\
&\!\!\!\!\leq \!\!\!\!&  I(W_{23},W_{43};Y_R^n,\mathcal{G}|W_{31},W_{32},W_{34})+\epsilon(n)\\
&\!\!\!\!= \!\!\!\!&  I(W_{23},W_{43};\mathcal{G}|W_{31},W_{32},W_{34})+I(W_{23},W_{43};Y_R^n|\mathcal{G},W_{31})+\epsilon(n)\\
&\!\!\!\!= \!\!\!\!&  I(W_{23},W_{43};Y_R^n|W_{31},\mathcal{G})+\epsilon(n)\\
&\!\!\!\!\leq \!\!\!\!&  I(W_{23},W_{43};Y_R^n|W_{21},W_{31},W_{41},\mathcal{G})+\epsilon(n)\label{eqn:Yob2_s}
\end{eqnarray}
where (\ref{eqn:Yob2_s}) follows from the fact that $I(A;B|C)\geq I(A;B)$ when $A$ is independent of $C$.

Next, consider the rate of $W_{24}$ desired at user 4:
\begin{eqnarray}
nR_{24}&\!\!\!\!\leq \!\!\!\!&  I(W_{24};Y_4^n|W_{41},W_{42},W_{43})+\epsilon(n)\\
&\!\!\!\!\leq \!\!\!\!&  I(W_{24};Y_R^n|W_{41},W_{42},W_{43})+\epsilon(n)\\
&\!\!\!\!\leq \!\!\!\!&  I(W_{24};Y_R^n,\mathcal{G}|W_{41},W_{42},W_{43})+\epsilon(n)\\
&\!\!\!\!= \!\!\!\!&  I(W_{24};\mathcal{G}|W_{41},W_{42},W_{43})+I(W_{24};Y_R^n|W_{41},W_{43},\mathcal{G})+\epsilon(n)\\
&\!\!\!\!= \!\!\!\!&  I(W_{24};Y_R^n|W_{41},W_{43},\mathcal{G})+\epsilon(n)\\
&\!\!\!\!\leq \!\!\!\!&  I(W_{24};Y_R^n|W_{21},W_{31},W_{41},W_{23},W_{43},\mathcal{G})+\epsilon(n)\ \ \ \label{eqn:Yob3_s}.
\end{eqnarray}

Finally, adding up (\ref{eqn:Yob1_s}), (\ref{eqn:Yob2_s}) and (\ref{eqn:Yob3_s}), we have:
\begin{eqnarray}
n(R_{21}+R_{31}+R_{41}+R_{23}+R_{43}+R_{24})&\!\!\!\!\leq \!\!\!\!&  I(W_{21},W_{31},W_{41},W_{23},W_{43},W_{24};Y_R^n|\mathcal{G})+\epsilon(n)\label{eqn:Yob4_chain}\\
&\!\!\!\!\leq \!\!\!\!& h(Y_R^n|\mathcal{G})+\epsilon(n)\\
&\!\!\!\!\leq \!\!\!\!& nN\log P+\epsilon(n)\label{eqn:Yob4_s}
\end{eqnarray}
where (\ref{eqn:Yob4_chain}) is obtained due to the mutual information chain rule, and (\ref{eqn:Yob4_s}) is obtained since the relay has a total of $N$ antennas. Then dividing $n\log P$ on both sides, and letting first $n\rightarrow\infty$ and then $P\rightarrow \infty$, we obtain the desired DoF upper bound
\begin{eqnarray}
6nR \leq nN\log P+\epsilon(n)\Longrightarrow d\leq N/6.
\end{eqnarray}

\subsection{$3/7\leq M/N \leq 1/2 \Longrightarrow d\leq N/7$}
Similar to last section, we still first provide a linear dimension counting approach to produce the DoF upper bound, and then translate it into the information theoretic statement.

Suppose that a genie provides to the relay $\mathcal{G}=\{W_{12},W_{13},W_{14},W_{42},W_{43}\}$. Now, let us count the number of dimensions contributed by the other 7 messages. Since user 1 can decode $W_{21},W_{31},W_{41}$ with its own 3 messages $W_{12},W_{13},W_{14}$ as side information, which are also provided to the relay by the genie, the relay is able to decode $W_{21},W_{31},W_{41}$ as well. Thus, these 3 messages contribute a total of $3d$ dimensions, which have to be linearly independent with the interfering dimensions contributed by the rest 4 messages $W_{23},W_{24},W_{32},W_{34}$. Note that $W_{23},W_{24}$ both originate from user 2 and $W_{32},W_{34}$ both originate from user 3, and these two users do not project a common intersection at the $N$-dimensional vector space at the relay because of $2M\leq N$. Thus, $W_{23},W_{24},W_{32},W_{34}$ contribute additional $4d$ dimensions. Therefore, the 7 messages $W_{21},W_{31},W_{41}$ and $W_{23},W_{24},W_{32},W_{34}$ contribute a total of $7d$ dimensions in the $N$-dimensional vector space at the relay, so that the symmetric DoF per message are upper bounded by $N/7$.

The linear dimension counting approach can be easily translated into the information theoretic statement. Let a genie provide to the relay the messages $\mathcal{G}=\{W_{12},W_{13},W_{14},W_{42},W_{43}\}$. Then we consider the sum rate of the 3 messages desired at user 1:
\begin{eqnarray}
n(R_{21}+R_{31}+R_{41})
&\!\!\!\!\leq \!\!\!\!&  I(W_{21},W_{31},W_{41};Y_1^n|W_{12},W_{13},W_{14})+\epsilon(n)\\
&\!\!\!\!\leq \!\!\!\!&  I(W_{21},W_{31},W_{41};Y_R^n|W_{12},W_{13},W_{14})+\epsilon(n)\\
&\!\!\!\!\leq \!\!\!\!&  I(W_{21},W_{31},W_{41};Y_R^n,\mathcal{G}|W_{12},W_{13},W_{14})+\epsilon(n)\\
&\!\!\!\!= \!\!\!\!&  I(W_{21},W_{31},W_{41};\mathcal{G}|W_{12},W_{13},W_{14})+I(W_{21},W_{31},W_{41};Y_R^n|\mathcal{G})+\epsilon(n)\\
&\!\!\!\!= \!\!\!\!& I(W_{21},W_{31},W_{41};Y_R^n|\mathcal{G})+\epsilon(n).\label{eqn:Y2ob1_s}
\end{eqnarray}

Next, consider the sum rate of the messages $W_{24}, W_{34}$ desired at user 4:
\begin{eqnarray}
n(R_{24}+R_{34})&\!\!\!\!\leq \!\!\!\!&  I(W_{24},W_{34};Y_4^n|W_{41},W_{42},W_{43})+\epsilon(n)\\
&\!\!\!\!\leq \!\!\!\!&  I(W_{24},W_{34};Y_R^n|W_{41},W_{42},W_{43})+\epsilon(n)\\
&\!\!\!\!\leq \!\!\!\!&  I(W_{24},W_{34};Y_R^n,\mathcal{G}|W_{41},W_{42},W_{43})+\epsilon(n)\\
&\!\!\!\!= \!\!\!\!&  I(W_{24},W_{34};\mathcal{G}|W_{41},W_{42},W_{43})+I(W_{24},W_{34};Y_R^n|W_{41},\mathcal{G})+\epsilon(n)\\
&\!\!\!\!= \!\!\!\!&  I(W_{24},W_{34};Y_R^n|W_{41},\mathcal{G})+\epsilon(n)\\
&\!\!\!\!\leq \!\!\!\!&  I(W_{24},W_{34};Y_R^n|W_{21},W_{31},W_{41},\mathcal{G})+\epsilon(n) \label{eqn:Y2ob2_s}.
\end{eqnarray}

Adding up (\ref{eqn:Y2ob1_s}), (\ref{eqn:Y2ob2_s}) we obtain:
\begin{eqnarray}
&&n(R_{21}+R_{31}+R_{41}+R_{24}+R_{34})\notag\\
&\!\!\!\!\leq \!\!\!\!&  I(W_{21},W_{31},W_{41},W_{24},W_{34};Y_R^n|\mathcal{G})+\epsilon(n)\\
&\!\!\!\!= \!\!\!\!& h(Y_R^n|\mathcal{G})-h(Y_R^n|\mathcal{G},W_{21},W_{31},W_{41},W_{24},W_{34})+\epsilon(n)\\
&\!\!\!\!= \!\!\!\!& h(Y_R^n|\mathcal{G})-h(Y_R^n|X_1^n,X_4^n,W_{21},W_{31},W_{24},W_{34})+\epsilon(n)\label{eqn:Y2ob3_recons}\\
&\!\!\!\!= \!\!\!\!& h(Y_R^n|\mathcal{G})-h(\tilde{X}_2^n,\tilde{X}_3^n|W_{21},W_{31},W_{24},W_{34})+n~\epsilon(\log P)+\epsilon(n)\label{eqn:Y2ob3_remov}\\
&\!\!\!\!= \!\!\!\!& h(Y_R^n|\mathcal{G})-H(W_{21},W_{23},W_{24},W_{31},W_{32},W_{34}|W_{21},W_{31},W_{24},W_{34})+n~\epsilon(\log P)+\epsilon(n)\label{eqn:Y2ob3_msg}\\
&\!\!\!\!= \!\!\!\!& h(Y_R^n|\mathcal{G})-H(W_{23},W_{32})+n~\epsilon(\log P)+\epsilon(n)\\
&\!\!\!\!\leq \!\!\!\!& nN\log P-n(R_{23}+R_{32})+n~\epsilon(\log P)+\epsilon(n)\label{eqn:Y2ob3_rearrange}
\end{eqnarray}
where (\ref{eqn:Y2ob3_recons}) is obtained because $X_1$ is an encoding function of the messages $W_{12},W_{13},W_{14}$, and $X_4$ is an encoding function of the messages $W_{41},W_{42},W_{43}$; (\ref{eqn:Y2ob3_remov}) follows from the fact that by subtracting the contributions of $X_1$ and $X_4$ from $Y_R$, i.e., $Y_R-{\bf H}_1X_1-{\bf H}_4X_4={\bf H}_2X_2+{\bf H}_3X_3+Z_R$, the relay only sees $N$ linear combinations of the signals $\tilde{X}_2$, $\tilde{X}_3$, i.e., $X_2$, $X_3$ sent from user 2 and user 3 subject to the noise. Since they both have $M$ antennas only and $2M\leq N$, the relay is able to recover the signals $X_2$ and $X_3$ subject to the noise distortion; (\ref{eqn:Y2ob3_msg}) is obtained because $X_2$ and $X_3$ are encoding functions of their 3 own messages, respectively.

Finally, by rearranging (\ref{eqn:Y2ob3_rearrange}) we have the following rate inequality:
\begin{eqnarray}
n(R_{21}+R_{31}+R_{41}+R_{24}+R_{34}+R_{23}+R_{32})\leq nN\log P+n~\epsilon(\log P)+\epsilon(n).
\end{eqnarray}
Dividing $n\log P$ on both sides, and letting first $n\rightarrow\infty$ and then $P\rightarrow \infty$, we obtain the desired DoF upper bound
\begin{eqnarray}
7nR \leq nN\log P+n~\epsilon(\log P)+\epsilon(n)\Longrightarrow d\leq N/7.
\end{eqnarray}

\subsection{$1/2\leq M/N \leq 7/12 \Longrightarrow d\leq 2M/7$}

The DoF converse proof for this regime is similar to that we show in last section because they both originate from the case of $M/N=1/2$ from the intuition of antenna dimension redundances, except that here we have $2M\geq N$ which implies that every two users project a $(2M-N)$-dimensional common intersection in the $N$-dimensional signal vector space at the relay. Observing the 7 messages $W_{21},W_{31},W_{41}$, $W_{23},W_{24},W_{32},W_{34}$ that we bound their sum rate in last section, we find that only $(W_{23},W_{32})$ are two pairwise messages. Since $2M\geq N$, the signals carrying these two pairwise messages can occupy a common subspace at the relay. Thus, from the DoF converse perspective, after a genie again provides $\{W_{12},W_{13},W_{14},W_{42},W_{43}\}$ to the relay, in order for the relay to decode the other 7 messages, a genie still needs to provide additional information associated with that common intersection projected from user 2 and user 3, so that $W_{23}$ and $W_{32}$ can be decoded as well. In order to do so, the additional information that the genie needs to provide is the signal from, e.g., user 3, which will be projected into the common intersection projected from user 2 and user 3 at the relay. We denote this $(2M-N)\times 1$ column signal vector from user 3 as $X_{3c}$, which is given by $X_{3c}=\left(({\bf H}_3^H {\bf H}_3)^{-1}{\bf H}_3^H[{\bf H}_2^c~{\bf H}_3^c]^c\right)^T X_3$, where ${\bf A}^c$ stands for the arbitrary set of basis of null space of the matrix ${\bf A}^T$. Then we directly provide the information theoretic DoF converse proof in the following.

Let a genie provide the information $\mathcal{G}=\{\mathcal{G}',X_{3c}^n\}$ where $\mathcal{G'}=\{W_{12},W_{13},W_{14},W_{42},W_{43}\}$ to the relay. Then we consider the sum rate of the 3 messages desired at user 1:
\begin{eqnarray}
&& n(R_{21}+R_{31}+R_{41})\notag\\
&\!\!\!\!\leq \!\!\!\!&  I(W_{21},W_{31},W_{41};Y_1^n|W_{12},W_{13},W_{14})+\epsilon(n)\\
&\!\!\!\!\leq \!\!\!\!&  I(W_{21},W_{31},W_{41};Y_R^n|W_{12},W_{13},W_{14})+\epsilon(n)\\
&\!\!\!\!\leq \!\!\!\!&  I(W_{21},W_{31},W_{41};Y_R^n,\mathcal{G}|W_{12},W_{13},W_{14})+\epsilon(n)\\
&\!\!\!\!= \!\!\!\!&  I(W_{21},W_{31},W_{41};\mathcal{G'}|W_{12},W_{13},W_{14})+I(W_{21},W_{31},W_{41};X_{3c}^n,Y_R^n|\mathcal{G'})+\epsilon(n)\\
&\!\!\!\!= \!\!\!\!& I(W_{21},W_{31},W_{41};X_{3c}^n,Y_R^n|\mathcal{G'})+\epsilon_n.\label{eqn:Y3ob1_s}
\end{eqnarray}

Next, consider the sum rate of the messages $W_{24}, W_{34}$ desired at user 4:
\begin{eqnarray}
n(R_{24}+R_{34})&\!\!\!\!\leq \!\!\!\!&  I(W_{24},W_{34};Y_4^n|W_{41},W_{42},W_{43})+\epsilon(n)\\
&\!\!\!\!\leq \!\!\!\!&  I(W_{24},W_{34};Y_R^n|W_{41},W_{42},W_{43})+\epsilon(n)\\
&\!\!\!\!\leq \!\!\!\!&  I(W_{24},W_{34};Y_R^n,\mathcal{G}|W_{41},W_{42},W_{43})+\epsilon(n)\\
&\!\!\!\!= \!\!\!\!&  I(W_{24},W_{34};\mathcal{G'}|W_{41},W_{42},W_{43})+I(W_{24},W_{34};X_{3c}^n,Y_R^n|W_{41},\mathcal{G}')+\epsilon(n)\\
&\!\!\!\!= \!\!\!\!&  I(W_{24},W_{34};X_{3c}^n,Y_R^n|W_{41},\mathcal{G'})+\epsilon(n)\\
&\!\!\!\!\leq \!\!\!\!&  I(W_{24},W_{34};X_{3c}^n,Y_R^n|W_{21},W_{31},W_{41},\mathcal{G'})+\epsilon(n) \label{eqn:Y3ob2_s}.
\end{eqnarray}

Adding up (\ref{eqn:Y3ob1_s}), (\ref{eqn:Y3ob2_s}) we obtain:
\begin{eqnarray}
&&n(R_{21}+R_{31}+R_{41}+R_{24}+R_{34})\notag\\
&\!\!\!\!\leq \!\!\!\!&  I(W_{21},W_{31},W_{41},W_{24},W_{34};X_{3c}^n,Y_R^n|\mathcal{G}')+\epsilon(n)\\
&\!\!\!\!= \!\!\!\!& h(X_{3c}^n,Y_R^n|\mathcal{G}')-h(X_{3c}^n,Y_R^n|\mathcal{G}',W_{21},W_{31},W_{41},W_{24},W_{34})+\epsilon(n)\\
&\!\!\!\!= \!\!\!\!& h(X_{3c}^n,Y_R^n|\mathcal{G}')-h(X_{3c}^n,Y_R^n|X_1^n,X_4^n,W_{21},W_{31},W_{24},W_{34})+\epsilon(n)\\
&\!\!\!\!= \!\!\!\!& h(X_{3c}^n,Y_R^n|\mathcal{G}')-h(\tilde{X}_2^n,\tilde{X}_3^n|W_{21},W_{31},W_{24},W_{34})+n~\epsilon(\log P)+\epsilon(n)\label{eqn:Y3ob3_remov}\\
&\!\!\!\!= \!\!\!\!& h(X_{3c}^n,Y_R^n|\mathcal{G}')-H(W_{23},W_{32})+n~\epsilon(\log P)+\epsilon(n)\\
&\!\!\!\!\leq \!\!\!\!& n(2M-N)\log P+nN\log P-n(R_{23}+R_{32})+n~\epsilon(\log P)+\epsilon(n)\label{eqn:Y3ob3_rearrange}
\end{eqnarray}
where (\ref{eqn:Y3ob3_remov}) is obtained because by removing the contributions of $X_1$ and $X_4$ from $Y_R$, the relay only sees $N$
linear combinations of the signals $\tilde{X}_2$, $\tilde{X}_3$, i.e., $X_2$, $X_3$ sent from user 2 and user 3 subject to the noise. Note that $X_{3c}$, the transmitted signal from user 3, is projected into that common subspace. Thus, from $(X_{3c},Y_R-{\bf H}_1X_1-{\bf H}_4X_4)$, the relay is able to reconstruct the signals $X_2$ and $X_3$ subject to the noise distortion.

Finally, by rearranging (\ref{eqn:Y3ob3_rearrange}) we have the following rate inequality:
\begin{eqnarray}
n(R_{21}+R_{31}+R_{41}+R_{24}+R_{34}+R_{23}+R_{32})\leq 2nM\log P+n~o(\log P)+\epsilon_n.
\end{eqnarray}
Then dividing $n\log P$ on both sides, and letting first $n\rightarrow\infty$ and then $P\rightarrow \infty$, we obtain the desired DoF upper bound
\begin{eqnarray}
7nR \leq 2nM\log P+n~o(\log P)+\epsilon_n\Longrightarrow d\leq 2M/7.
\end{eqnarray}

\section{DoF Achievability: Proof of Lemma 2}\label{sec:y_ib}

Following from the antenna dimension redundancies argument mentioned in Section \ref{sec:results}, and as what we illustrated in \cite{Wang_Gou_Jafar_3userMxN}, we again only need to first present the achievability schemes at the two points of $M/N=3/7$ and $M/N=7/12$, and then extend the DoF achievability to other regimes by the use of spatial extensions. Basically, the DoF achievability at $M/N=7/12$, reported in \cite{Yuan_xjtu}, is not very challenging due to the fact that only one-to-one alignment is sufficient. To see this, consider $(M,N)=(7,12)$ where each message carries 2 DoF. Since $2M-N=2$, every two users project a 2-dimensional common intersection at the relay. Thus, each user sends each message so that every two pairwise messages occupy only 2 dimensions at the relay. Since there are 6 pair of messages, the total number of dimensions occupied by all the 6 pairs is given by $2\times 6=12$. That is, one-to-one alignment is sufficient in this case. For the regime $M/N\geq 1/2$, we can also use the same achievable scheme, in the sense of spatial extensions.

In contrast with $M/N=7/12$, the DoF achievability at $M/N=3/7$ is much more challenging. In this section, we are primarily interested in this case, and the other regimes $M/N<3/7$ and $3/7<M/N\leq 1/2$ can then be covered by the same achievable scheme by using spatial extensions. As a remark, we emphasize here again that our DoF achievability only relies on linear beamforming schemes. Due to the reciprocity of linear schemes, and following from the prior work such as \cite{Lee_Lim_MIMOY, Wang_Jafar_2way}, once we finish the transmitting beamforming design at each user and the receiving beamforming design at the relay in the first phase, then transmitting beamforming at the relay and receiving beamforming at each user if the second phase can be automatically determined by using a reciprocal approach. By doing so, each user finally sees an equivalent interference-free single-user MIMO channel for each desired message. Thus, the key of designing the DoF achievability is the transmission design for the first phase, i.e., from each user to the relay.

\subsection*{Principle of the beamforming design in the first phase}

As mentioned in Section \ref{sec:y_ob}, the received signal at each user only depends on the transmitted signal at the relay. Since each user is able to decode its desired messages, with its own transmitted messages as side information, if that side information is available at the relay, then the relay is able to decode those messages as well. Thus, from the DoF achievability perspective, it does not affect whether the signals carrying the pairwise messages $W_{ij},W_{ji}$ align or not at the relay, because $W_{ij}$ is side information at user $i$ who demands $W_{ji}$, and $W_{ji}$ is side information at user $j$ who demands $W_{ij}$. Thus, the principle to design an achievability scheme is that \emph{for every pairwise messages $(W_{ij},W_{ji})$, the signals carrying all the other messages occupy only $N-d$ dimensions at the relay, and the $d$ dimensions at the relay accommodating each message in the pair $W_{ij},W_{ji}$ are linearly independent with those $N-d$ dimensions, so that the relay is able to resolve $d$ linearly independent combinations of the signals carrying $W_{ij},W_{ji}$ only, subject to the noise.}

\subsection{$(M,N)=(3,7)\Longrightarrow d=1$}

For this case, Lemma 2 implies that each message can achieve $d=1$ DoF, i.e., each user can achieve a total of $3$ DoF which are also the single-user DoF upper bound.

\subsubsection{From users to the relay}

In the first phase, user $k$ encodes each symbol $u_{kj}$ using a $3\times 1$ beamforming vector $V_{kj},j\in\mathcal{K}\setminus\{k\}$, and the transmitted signal vector of user $k$, denoted as $X_k$, can be written as
\begin{eqnarray}
X_k = \sum_{j\in\mathcal{K}\setminus\{k\}} V_{kj} u_{kj}.
\end{eqnarray}
Now, let us consider the received signals at the relay node. Note that there are a total of 12 symbols, comprising 6 pair of messages $(u_{kj},u_{jk})$. As mentioned in the principle of the beamforming design in the first phase, \emph{our goal is that for every two pairwise messages, the other 5 pairs of symbols, i.e., the other 10 symbols, span $7-1=6$ dimensions only, and the signal vector carrying each symbol in that pair is linearly independent with those 6 dimensions, so that a linear combination of every two pairwise can be resolved at the relay.} By doing so, in the second phase, the relay can precode the 6 linear combinations of pairwise symbols using beamforming, so that every user finally only sees 3 linear combinations, each of which is associated with one of its own symbols and one of its desired symbols. With this approach, let us count the total number of alignment equations that we need. For every two pairwise symbols, we need to accommodate the other 10 symbols into a 6-dimensional subspace at the relay. That is, we need to align 4 symbols into the subspace spanned by the other symbols. Since there are 6 pairs, the total number of alignment equations is given by $4\times 6=24$. However, with such a counting approach, the associated 24 alignment equations could be linearly dependent with each other, and it is even not clear how to explicitly design each alignment equation. Thus, such a counting approach may not help solve this problem very much.

Instead, let us let us consider necessary vector alignment from the user perspective. Regarding every user, except for its 3 desired symbols and its 3 own symbols, the other $12-3-3=6$ interfering symbols constitute interference. Thus, we need to ensure that at the relay, for each user, the 3 desired symbols occupy a 3-dimensional subspace, which has only null intersection with the subspace spanned by the 6 interfering symbols. For example, consider user 1 who desires the symbols $u_{21}$, $u_{31}$ and $u_{41}$, and its own transmitted symbols are $u_{12}$, $u_{13}$ and $u_{14}$. Except for these 6 symbols, the other $u_{23}$, $u_{32}$, $u_{24}$, $u_{42}$, $u_{34}$ and $u_{43}$ are 6 interfering symbols. In the $7$-dimensional space at the relay, in order to protect a $3$-dimensional subspace for the 3 desired symbols, the 6 interfering symbols sent from from user 2, user 3 and user 4 can only occupy $7-3=4$ dimensions. That is, we need to align at least 2 of those 6 symbols into the subspace spanned by the other 4 symbols at the relay, implying that we need 2 alignment equations. Following from a symmetric analysis, if we consider user 2, user 3 and user 4, individually, we need a total of 8 alignment equations, 2 for each user. Note that this is only a necessary requirement, and our achievability scheme finally must meet the goal we emphasized above.

So far, we demonstrate that we need at least 8 alignment equations. However, it is still quite challenging to design an achievable scheme for two coupled reasons.
\begin{itemize}
\item{} First, it is quite challenging to explicitly design the two alignment equations for each user. Consider the 6 interfering symbols at the relay regarding user 1 as an example. Note that $u_{23}$, $u_{24}$ both originate from user 2, $u_{32}$, $u_{34}$ both originate from user 3, and $u_{42}$, $u_{43}$ both originate from user 4. Since $2M<N$, every two users do not project a common intersection at the relay. Thus, finding the explicit two alignment equation by identifying which symbol is aligned into the subspace spanned by which symbols, is essentially quite important and very challenging.

\item{} Second, as we will explain later, in fact, the 8 alignment equations are still linearly dependent, i.e., some equations can be linearly represented by the others. Thus, which alignment equations are redundant among the 8 alignment equations is nontrivial. To see this, let us count the number of nulling equations and variables. Since we need to design 12 $3\times 1$ beamforming vectors, there are a total of 36 variables we need to determine their values. Also, each alignment equation at the relay contributes $N=7$ nulling equations. If all the 8 alignment equations, i.e., $8N=56$ nulling equations are linearly independent, then linear algebra implies that the beamforming vectors have to be zero. In order to have a non-zero solution, one necessary condition is that we must have fewer linearly independent nulling equations than the number of variables. Thus, among the 56 nulling equations, at most there are $36-1=35$ linearly independent nulling equations. That is, there are at least 21 redundant nulling equations, or equivalently 3 redundant alignment equations. However, the question which 3 alignment equations are redundant is nontrivial.
\end{itemize}

As mentioned above, since directly designing each alignment equation might be quite difficult, let us first consider the signal vectors at the relay carrying each symbol. We denote $F_i,i=1,2,\cdots,7$ as 7 linearly independent $7\times 1$ column vectors. Suppose that the three vectors carrying the 3 symbols of user 1, i.e., $V_{12}$, $V_{13}$ and $V_{14}$, arrive at the relay along the following vectors:
\begin{eqnarray}
{\bf H}_1V_{12}&\!\!\!\!=\!\!\!\!&-(F_1+F_4+F_7),\label{eqn:Yu1_2}\\
{\bf H}_1V_{13}&\!\!\!\!=\!\!\!\!&-(F_2+F_4+F_7),\label{eqn:Yu1_3}\\
{\bf H}_1V_{14}&\!\!\!\!=\!\!\!\!&-(F_4+F_5+F_7).\label{eqn:Yu1_4}
\end{eqnarray}
Also, we suppose the 3 symbols of user 2 arrive at the relay along the vectors:
\begin{eqnarray}
{\bf H}_2V_{21}&\!\!\!\!=\!\!\!\!&F_1,\label{eqn:Yu2_1}\\
{\bf H}_2V_{23}&\!\!\!\!=\!\!\!\!&-(F_3+F_4+F_7),\label{eqn:Yu2_3}\\
{\bf H}_2V_{24}&\!\!\!\!=\!\!\!\!&-(F_4+F_6+F_7),\label{eqn:Yu2_4}
\end{eqnarray}
the 3 symbols of user 3 arrive at the relay along the vectors:
\begin{eqnarray}
{\bf H}_3V_{31}&\!\!\!\!=\!\!\!\!&F_2,\label{eqn:Yu3_1}\\
{\bf H}_3V_{32}&\!\!\!\!=\!\!\!\!&F_3,\label{eqn:Yu3_2}\\
{\bf H}_3V_{34}&\!\!\!\!=\!\!\!\!&F_4,\label{eqn:Yu3_4}
\end{eqnarray}
and the 3 symbols of user 4 at the relay along the vectors:
\begin{eqnarray}
{\bf H}_4V_{41}&\!\!\!\!=\!\!\!\!&F_5,\label{eqn:Yu4_1}\\
{\bf H}_4V_{42}&\!\!\!\!=\!\!\!\!&F_6,\label{eqn:Yu4_2}\\
{\bf H}_4V_{43}&\!\!\!\!=\!\!\!\!&F_7.\label{eqn:Yu4_3}
\end{eqnarray}

As we will show the linear independencies among vectors $F_i,i=1,2,\cdots,7$ later, let us first examine at the relay, for every symbols pair, the other 5 pairs, i.e., 10 symbols, only span a $6$-dimensional subspace, and the vector carrying each symbol in that pair does not align into that $6$-dimensional subspace spanned by the other 5 pairs. There are 6 symbols pairs, $(u_{12},u_{21})$, $(u_{13},u_{31})$, $(u_{14},u_{41})$, $(u_{23},u_{32})$, $(u_{24},u_{42})$ and $(u_{34},u_{43})$. Let us examine each symbols pair individually.
\begin{itemize}
\item{} For the pair of symbols $(u_{12},u_{21})$ at the relay, the vectors carrying the other 10 symbols $u_{13}$, $u_{14}$, $u_{31}$, $u_{41}$, $u_{23}$, $u_{24}$, $u_{32}$, $u_{34}$, $u_{42}$ and $u_{43}$ at the relay, are given by $-(F_2+F_4+F_7)$, $-(F_4+F_5+F_7)$, $F_2$, $F_5$, $-(F_3+F_4+F_7)$, $-(F_4+F_6+F_7)$, $F_3$, $F_4$, $F_6$ and $F_7$, as shown in (\ref{eqn:Yu1_3}), (\ref{eqn:Yu1_4}), (\ref{eqn:Yu3_1}), (\ref{eqn:Yu4_1}), (\ref{eqn:Yu2_3}), (\ref{eqn:Yu2_4}), (\ref{eqn:Yu3_2}), (\ref{eqn:Yu3_4}), (\ref{eqn:Yu4_2}) and (\ref{eqn:Yu4_3}), respectively. It can be seen that all these 10 symbols lie in a $6$-dimensional subspace spanned by the 6 linearly independent basis $F_2$, $F_3$, $F_4$, $F_5$, $F_6$ and $F_7$. In addition, the desired symbol $u_{12}$ for user 2, and the desired symbol $u_{21}$ for user 1, arrive at the relay along the vectors $-(F_1+F_4+F_7)$ and $F_1$ respectively, each of which does not lie in that $6$-dimensional subspace spanned by $F_2$, $F_3$, $F_4$, $F_5$, $F_6$, $F_7$.

\item{} For the pair $(u_{13},u_{31})$ at the relay, the vectors of the other 10 symbols $u_{12}$, $u_{14}$, $u_{21}$, $u_{41}$, $u_{23}$, $u_{24}$, $u_{32}$, $u_{34}$, $u_{42}$ and $u_{43}$, are given by $-(F_1+F_4+F_7)$, $-(F_4+F_5+F_7)$, $F_1$, $F_5$, $-(F_3+F_4+F_7)$, $-(F_4+F_6+F_7)$, $F_3$, $F_4$, $F_6$ and $F_7$, respectively. Thus, all these 10 symbols lie in a $6$-dimensional subspace spanned by the basis $F_1$, $F_3$, $F_4$, $F_5$, $F_6$ and $F_7$. In addition, the desired symbol $u_{13}$ for user 3, and the desired symbol $u_{31}$ for user 1, arrive at the relay along the vectors $-(F_2+F_4+F_7)$ and $F_2$ respectively, each of which is linearly independent with the 6 basis $F_1$, $F_3$, $F_4$, $F_5$, $F_6$, $F_7$.

\item{} For the pair $(u_{14},u_{41})$ at the relay, the vectors of the other 10 symbols $u_{12}$, $u_{13}$, $u_{21}$, $u_{31}$, $u_{23}$, $u_{24}$, $u_{32}$, $u_{34}$, $u_{42}$ and $u_{43}$, are given by $-(F_1+F_4+F_7)$, $-(F_2+F_4+F_7)$, $F_1$, $F_2$, $-(F_3+F_4+F_7)$, $-(F_4+F_6+F_7)$, $F_3$, $F_4$, $F_6$ and $F_7$, respectively. Thus, all these 10 symbols lie in a $6$-dimensional subspace spanned by the basis $F_1$, $F_2$, $F_3$, $F_4$, $F_6$ and $F_7$. In addition, the desired symbol $u_{14}$ for user 4, and the desired symbol $u_{41}$ for user 1, arrive at the relay along the vectors $-(F_4+F_5+F_7)$ and $F_5$ respectively, each of which is linearly independent with the 6 basis $F_1$, $F_2$, $F_3$, $F_4$, $F_6$, $F_7$.

\item{} For the pair $(u_{23},u_{32})$ at the relay, the vectors of the other 10 symbols $u_{12}$, $u_{13}$, $u_{21}$, $u_{31}$, $u_{14}$, $u_{24}$, $u_{41}$, $u_{34}$, $u_{42}$ and $u_{43}$, are given by $-(F_1+F_4+F_7)$, $-(F_2+F_4+F_7)$, $F_1$, $F_2$, $-(F_4+F_5+F_7)$, $-(F_4+F_6+F_7)$, $F_5$, $F_4$, $F_6$ and $F_7$, respectively. Thus, all these 10 symbols lie in a $6$-dimensional subspace spanned by the basis $F_1$, $F_2$, $F_4$, $F_5$, $F_6$ and $F_7$. In addition, the desired symbol $u_{23}$ for user 3, and the desired symbol $u_{32}$ for user 2, arrive at the relay along the vectors $-(F_3+F_4+F_7)$ and $F_3$ respectively, each of which is linearly independent with the 6 basis $F_1$, $F_2$, $F_4$, $F_5$, $F_6$, $F_7$.

\item{} For the pair $(u_{24},u_{42})$ at the relay, the vectors of the other 10 symbols $u_{12}$, $u_{13}$, $u_{21}$, $u_{31}$, $u_{14}$, $u_{23}$, $u_{41}$, $u_{34}$, $u_{32}$ and $u_{43}$, are given by $-(F_1+F_4+F_7)$, $-(F_2+F_4+F_7)$, $F_1$, $F_2$, $-(F_4+F_5+F_7)$, $-(F_3+F_4+F_7)$, $F_5$, $F_4$, $F_3$ and $F_7$, respectively. Thus, all these 10 symbols lie in a $6$-dimensional subspace spanned by the basis $F_1$, $F_2$, $F_3$, $F_4$, $F_5$ and $F_7$. In addition, the desired symbol $u_{24}$ for user 4, and the desired symbol $u_{42}$ for user 2, arrive at the relay along the vectors $-(F_4+F_6+F_7)$ and $F_6$ respectively, each of which is linearly independent with the 6 basis $F_1$, $F_2$, $F_3$, $F_4$, $F_5$, $F_7$.

\item{} For the pair $(u_{34},u_{43})$ at the relay, the vectors of the other 10 symbols $u_{12}$, $u_{13}$, $u_{21}$, $u_{31}$, $u_{14}$, $u_{23}$, $u_{41}$, $u_{24}$, $u_{32}$ and $u_{42}$, are given by $-(F_1+F_4+F_7)$, $-(F_2+F_4+F_7)$, $F_1$, $F_2$, $-(F_4+F_5+F_7)$, $-(F_3+F_4+F_7)$, $F_5$, $-(F_4+F_6+F_7)$, $F_3$ and $F_6$, respectively. Thus, all these 10 symbols lie in a $6$-dimensional subspace spanned by the basis $F_1$, $F_2$, $F_3$, $F_5$ $F_6$, and $(F_4+F_7)$. In addition, the desired symbol $u_{34}$ for user 4, and the desired symbol $u_{43}$ for user 3, arrive at the relay along the vectors $F_4$ and $F_7$ respectively, each of which is linearly independent with the 6 basis $F_1$, $F_2$, $F_3$, $F_5$ $F_6$, $(F_4+F_7)$.
\end{itemize}
So far, it can be seen that for each pair of symbols, the other 10 symbols only occupy 6 dimensions, which are linearly independent with the signal vector carrying each symbol in the pair of interest.

What remains to be shown is how to find out the 7 linearly independent column vectors $F_i,i=1,2,\cdots,7$ given the matrices ${\bf H}_k,k\in\mathcal{K}$. Let us look into the 12 vectors on the right-hand-side of equations from (\ref{eqn:Yu1_2}) to (\ref{eqn:Yu4_3}). It turns out that only the 5 vectors in (\ref{eqn:Yu1_2}), (\ref{eqn:Yu1_3}), (\ref{eqn:Yu1_4}), (\ref{eqn:Yu2_3}), (\ref{eqn:Yu2_4}) are linear combinations of the other 7 vectors. Thus, if we rewrite all equations from (\ref{eqn:Yu1_2}) to (\ref{eqn:Yu4_3}) by eliminating all $F_i,i=1,2,\cdots,7$, we obtain the following 5 alignment equations:
\begin{eqnarray}
{\bf H}_1V_{12}+{\bf H}_2V_{21}+{\bf H}_3V_{34}+{\bf H}_4V_{43}&\!\!\!\!=\!\!\!\!&0,\label{eqn:Yalign1}\\
{\bf H}_1V_{13}+{\bf H}_3V_{31}+{\bf H}_3V_{34}+{\bf H}_4V_{43}&\!\!\!\!=\!\!\!\!&0,\\
{\bf H}_1V_{14}+{\bf H}_3V_{34}+{\bf H}_4V_{41}+{\bf H}_4V_{43}&\!\!\!\!=\!\!\!\!&0,\\
{\bf H}_2V_{23}+{\bf H}_3V_{32}+{\bf H}_3V_{34}+{\bf H}_4V_{43}&\!\!\!\!=\!\!\!\!&0,\\
{\bf H}_2V_{24}+{\bf H}_3V_{34}+{\bf H}_4V_{42}+{\bf H}_4V_{43}&\!\!\!\!=\!\!\!\!&0.\label{eqn:Yalign5}
\end{eqnarray}
As shown above, the signal alignment happens among the subspaces projected from different user nodes. Such an operation is referred to as inter-user signal subspace alignment, which based on one-to-many alignment is essentially the key to design the DoF achievability in this work. So far, we settle the two challenges that we mention earlier, i.e., removing redundant alignment equations from the 8 alignment conditions, and specifying each alignment equation. Next, let us rewrite the 5 alignment equations from (\ref{eqn:Yalign1}) to (\ref{eqn:Yalign5}) into a compact matrix form as follows:
\begin{eqnarray}
\underbrace{\left[\begin{array}{cccccccccccc}
{\bf H}_1&{\bf O}&{\bf O}&{\bf H}_2&{\bf O}&{\bf O}&{\bf O}&{\bf O}&{\bf H}_3&{\bf O}&{\bf O}&{\bf H}_4\\
{\bf O}&{\bf H}_1&{\bf O}&{\bf O}&{\bf O}&{\bf O}&{\bf H}_3&{\bf O}&{\bf H}_3&{\bf O}&{\bf O}&{\bf H}_4\\
{\bf O}&{\bf O}&{\bf H}_1&{\bf O}&{\bf O}&{\bf O}&{\bf O}&{\bf O}&{\bf H}_3&{\bf H}_4&{\bf O}&{\bf H}_4\\
{\bf O}&{\bf O}&{\bf O}&{\bf O}&{\bf H}_2&{\bf O}&{\bf O}&{\bf H}_3&{\bf H}_3&{\bf O}&{\bf O}&{\bf H}_4\\
{\bf O}&{\bf O}&{\bf O}&{\bf O}&{\bf O}&{\bf H}_2&{\bf O}&{\bf O}&{\bf H}_3&{\bf O}&{\bf H}_4&{\bf H}_4
\end{array}\right]}_{\triangleq {\bf H}_{35\times 36}}\!\!
\left[\begin{array}{c}V_{12}\\V_{13}\\ \vdots \\ V_{43}\end{array}\right]=0.\label{eqn:Yalign_c}
\end{eqnarray}
Since ${\bf H}_k,k\in\mathcal{K}$ are generic, it is not difficult to verify the $35\times 36$ matrix ${\bf H}$ in (\ref{eqn:Yalign_c}) has full rank, by picking a special set of ${\bf H}_k,k\in\mathcal{K}$ and compute $\det({\bf H}{\bf H}^H)\neq 0$\footnote{This approach has been widely used in information theory to study linear independencies among vectors of a matrix, such as \cite{Wang_Gou_Jafar_3userMxN}.}. For example, we pick the following set of ${\bf H}_k,k\in\mathcal{K}$:
\begin{eqnarray}
{\bf H}_1\!=\!\left[\begin{array}{ccc}1&0&0\\0&1&0\\0&0&1\\0&0&0\\0&0&0\\0&0&0\\0&0&0\end{array}\right],~~~{\bf H}_2\!=\!\left[\begin{array}{ccc}0&0&0\\0&0&0\\0&0&0\\1&0&0\\0&1&0\\0&0&1\\0&0&0\end{array}\right],~~~{\bf H}_3\!=\!\left[\begin{array}{ccc}0&1&0\\0&0&1\\0&0&0\\0&1&0\\0&0&1\\0&0&0\\1&0&0\end{array}\right],~~~{\bf H}_4\!=\!\left[\begin{array}{ccc}0&1&0\\0&1&1\\1&0&0\\0&1&0\\0&0&1\\1&0&1\\0&1&1\end{array}\right],\label{eqn:Yspecial}
\end{eqnarray}
and then it is easy to verify that ${\bf H}$ has full rank by computing $\det({\bf H}{\bf H}^H)\neq 0$. As a consequence, from (\ref{eqn:Yalign_c}), we obtain the beamforming vectors of each user as follows:
\begin{eqnarray}
V=({\bf I}_{36}-{\bf H}^H({\bf H}{\bf H}^H)^{-1}{\bf H})\det({\bf H}{\bf H}^H)Q_{36\times 1},\ \ \ \ \ \ \ \ \label{eqn:YV}\\
\begin{array}{lll}
V_{12}=V(1:3),&V_{13}=V(4:6),&V_{14}=V(7:9),\\
V_{21}=V(10:12),&V_{23}=V(13:15),&V_{24}=V(16:18),\\
V_{31}=V(19:21),&V_{32}=V(22:24),&V_{34}=V(25:27),\\
V_{41}=V(28:30),&V_{42}=V(31:33),&V_{43}=V(34:36),
\end{array}\!\!\!\!
\end{eqnarray}
where $Q_{36\times 1}$ is a randomly picked $36\times 1$ vector.

Finally, it suffices to only prove that the 7 column vectors $F_i,i=1,\cdots,7$ are linearly independent, or equivalently to show that the following matrix has full rank:
\begin{eqnarray*}
{\bf G}\triangleq [F_1,F_2,F_3,F_4,F_5,F_6,F_7]=[{\bf H}_2V_{21},{\bf H}_3V_{31},{\bf H}_3V_{32},{\bf H}_3V_{34},{\bf H}_4V_{41},{\bf H}_4V_{42},{\bf H}_4V_{43}].
\end{eqnarray*}
Since each entry of ${\bf G}$ is a polynomial of the entries of ${\bf H}_k,k\in\mathcal{K}$, to show ${\bf G}$ has full rank, it suffices to show $\det({\bf G})\neq 0$ via picking a special set of ${\bf H}_k,k\in\mathcal{K}$. Again, we pick the special matrices in (\ref{eqn:YV}), and the resulting beamforming vectors of each user are given by:
\begin{eqnarray}
\!\!\!\!\!\!\!\!&&\!\!\!\!\!\!\!\![V_{12},V_{13},V_{14}]\!=\!\alpha\left[\begin{array}{rrr}1&0&0\\0&1&0\\0&0&-1\end{array}\right],~~~~
[V_{21},V_{23},V_{24}]\!=\!\alpha\left[\begin{array}{rrr}1&0&0\\-1&-1&0\\0&0&1\end{array}\right],\\
\!\!\!\!\!\!\!\!&&\!\!\!\!\!\!\!\![V_{31},V_{32},V_{34}]\!=\!\alpha\left[\begin{array}{rrr}0&0&1\\1&1&0\\-1&0&1\end{array}\right],~~~~
[V_{41},V_{42},V_{43}]\!=\!\alpha\left[\begin{array}{rrr}1&0&0\\1&1&-1\\-1&-1&0\end{array}\right]\!\!,\notag
\end{eqnarray}
where $\alpha$ is a non-zero scalar to satisfy the power constraint. Then it is easy to verify $\det({\bf G})\neq 0$, i.e., the matrix ${\bf G}$ has full rank.

So far, we finish the design of the beamforming vectors design in the first phase, from each user to the relay.

\subsubsection{From the relay to users}
In the first phase, the relay is able to resolve a clean linear combination of every pair of symbols $(u_{ij},u_{ji})$, by projecting the received signal into the nullspace of the 6-dimensional subspace spanned by the other 10 symbols. By doing so, the relay obtains 6 linear combinations associated with every two pairwise symbols, subject to the Gaussian noise:
\begin{eqnarray}
s_{ij}=\beta_{ij}u_{ij}+\beta_{ji}'u_{ji}+z_{ij},~~~~i,j\in\mathcal{K},i\neq j
\end{eqnarray}
where the linear combination coefficients $\beta_{ij},\beta_{ij}'$ depend on the channel matrices ${\bf H}_k,k\in\mathcal{K}$ only, and $z_{ij}$ is the AWGN with bounded variance, which does not depend on the power $P$.

In the second phase, each user follows from a reciprocal approach to design its receiving beamforming matrix. Specifically, by replacing ${\bf H}_k$ in the first phase with $\bar{{\bf H}}_k^T$, and by using the same design as in the first phase, the transmitting beamforming vector for each symbol in the first phase, is the receiving beamforming vector for its pairwise symbol in the second phase. Recall that in the first phase, at the relay side, for each pair, the other 10 symbols only project a $6$-dimensional subspace at the relay. Thus, in the second phase, the relay sends each linear combination $s_{ij}$ to both user $i$ and user $j$ with the beamforming vector which is orthogonal to the 6-dimensional subspace projected back from the subspaces of all users accommodating the those 10 symbols. By doing so, each user $j$ finally only sees 3 linear combinations $s_{ij},~i\neq j$, subject to the Gaussian noise. Note that each linear combination is associated with $u_{ij},u_{ji}$ only subject to the noise. Since we are primarily interested in the DoF characterization, i.e., the noise term can be neglected from the DoF perspective, by subtracting the signal carrying its own symbol $u_{ji}$, user $j$ sees an equivalent single-user point-to-point channel $s_{ij}-\beta_{ji}'u_{ji}=\beta_{ij}u_{ij}+z_{ij}$, free of interference. Thus, user $j$ is able to decode its desired symbol ${u}_{ij}$, so that each message can achieve 1 DoF.

\subsection{General Cases in the Regime $M/N\leq 1/2$}

For general $M/N=3/7$ cases, i.e., $(M,N)=(3\beta,7\beta)$ but $\beta\in\mathbb{Z}^+,~\beta>1$, we can still use the same achievable scheme that we show in last section, as long as each symbol is a $\beta$-dimensional symbol, i.e., each symbol/message carries $\beta$ DoF.

For the other cases in the regime $M/N\leq 1/2$, similar to \cite{Wang_Gou_Jafar_3userMxN}, we establish the DoF achievability in the spatial extension sense. In particular, if $M/N<3/7$, we scale the number of antennas at each node by $3$, to obtain a $(M',N')=(3M,3N)$ network. Since $3N>7M$, we reduce the number of antennas at the relay from $3N$ to $7M$. For this reduced network, we again can use the same approach that we show in last section to achieve $M$ DoF per message, so that the DoF per message normalized to spatial extensions are given by $M/3$, as shown in Theorem \ref{thm:y}. On the other hand, if $M/N>3/7$, we scale the number of antennas at each node by $7$, to obtain a $(M',N')=(7M,7N)$ network. Since $7M>3N$, we reduce the number of antennas at each user from $7M$ to $3N$. For this reduced network, we again use the same approach that we show in last section to achieve $N$ DoF per message, so that the DoF per message normalized to spatial extensions are given by $N/7$, as shown in Theorem \ref{thm:y}.

\section{Conclusion}

In this paper, we characterize the degrees of freedom (DoF) of two kinds multi-way relay MIMO interference networks, the 4-user relay MIMO $Y$ channel, and the two-way relay MIMO $X$ channel. To establish the DoF converse, we begin with the more intuitive linear dimension counting approach to produce the linear DoF upper bounds, and then translate them into the information theoretic upper bounds for each case. The information theoretic DoF upper bounds are facilitated by translating the multi-way two-hop channel to a
one-hop Multiple Access Channel with additional decoding constraints at the relay. Then by providing the relay enough genie information, the relay is able to decode all messages. To establish the DoF achievability, we propose linear beamforming schemes, in the sense of spatially-normalized extension in general, to show that the information theoretically optimal DoF can be achieved.

Several interesting observations follow as a byproduct of our analysis. First, we precisely identify settings with redundant dimensions at each user node, the relay node, both or neither. This observation, first identified in the one-way MIMO interference channel
in \cite{Wang_Gou_Jafar_3userMxN}, appears to exist in the multi-way relay MIMO interference networks as well. Second, our results in this paper also shed lights on the feasibility of linear interference alignment which has only been studied for one-way one-hop MIMO interference networks before. In this work, we show that many interesting observations such as improper is infeasible, many proper systems are infeasible, available for one-way one-hop MIMO interference channels \cite{Wang_Gou_Jafar_3userMxN}, are verified to exist in multi-way relay networks that we study in this work as well. Finally, while we find that the DoF counting bound also serves the upper bound for linear beamforming schemes, we are quite of interest in the question whether the DoF decomposition bound exists in the multi-way relay communication networks. The DoF decomposition is already well known for several one-way interference channels such as \cite{Wang_Sun_Jafar_4userMxN} and $X$ channels \cite{Sun_Geng_Gou_Jafar}. The relationship between the DoF counting bound and the DoF decomposition bound motivates many recent interesting work for the one-way channels such as \cite{Wang_Gou_Jafar_3userMxN, Wang_Sun_Jafar_4userMxN, Sun_Geng_Gou_Jafar, feasibility_spain}, it is not clear if we have the same observation or fundamental principles behind the results for multi-way communication networks. The answer to this question will be helpful to characterize the DoF of general multi-way relay MIMO interference networks, particularly when the number of users or the number of messages increase, such as the $K>4$ user relay MIMO $Y$ channel and the $K>3$ user pair two-way relay MIMO interference channel. All these research avenues would be of interest in our future work.

\newpage

\appendix
\section*{Appendix}
\section{Development of the DoF Counting Bound}\label{app:count}

As introduced earlier, the key to the DoF achievability schemes is the design of beamforming at each user in the first phase, to guarantee that a $d$-dimensional clean subspace can be protected for each pair of messages, and then the transmission scheme in the second phase can be automatically determined in a reciprocal manner. For developing the DoF counting bound for the two networks we study in this work, we also follow from this principle, and mimic the counting approach explicitly illustrated in \cite{Cenk_Gou_Jafar} for the one-way one-hop MIMO interference channel.

\subsection{All Unicast Setting: 4-User Relay MIMO $Y$ Channel}\label{app:county}

Let us first count the number of variables. At the user side, each user encodes the $d$ independent symbols carrying each message with an $M\times d$ beamforming matrix. Thus, the beamforming matrix corresponding to each message contributes $(M-d)d$ independent variables. Since there are a total of $12$ messages, the number of variables contributed by the beamforming matrices of all the $12$ messages is given by $12(M-d)d$. At the relay node, to protect a $d$-dimensional clean subspace for each pair of messages, the relay employs an $N\times d$ receiving beamforming matrix. Since the 12 messages comprise a total of $6$ pairs, the number of variables contributed by the receiving beamforming matrices at the relay node is $6(N-d)d$. Thus, the total number of variables is given by $12(M-d)d+6(N-d)d$.

Next, we count the number of nulling equations. For each pair of messages, when the relay employs an $N\times d$ receiving beamforming matrix to protect a $d$-dimensional clean subspace for that pair, the relay essentially zero forces all the signals carrying the other $10$ messages, each encoded to $d$ independent symbols. For every pair of messages, the total number of nulling equations is thus given by $10d^2$. Hence, the number of nulling equations is $60d^2$.

Following from the intuition explained in \cite{Cenk_Gou_Jafar}, we define a system is \emph{proper} if the number of variables is no fewer than the number of nulling equations. Thus, we obtain the following inequality to produce the DoF counting bound:
\begin{eqnarray}
12(M-d)d+6(N-d)d\geq 60d^2 \Longrightarrow d\leq \frac{2M+N}{13}.
\end{eqnarray}

\subsection{Multiple Unicast Setting: Two-Way Relay MIMO $X$ Channel}\label{app:countx}

Similarly, for the multiple unicast setting, at the user side, each user encodes the $d$ independent symbols carrying each message with an $M\times d$ beamforming matrix. Since there are a total of $8$ messages in this network, the number of variables contributed by the beamforming matrices of all the $8$ messages is $8(M-d)d$. At the relay node, to protect a $d$-dimensional clean subspace for each pair of messages, the relay node employs an $N\times d$ receiving beamforming matrix. Since the 8 messages comprise a total of $4$ pairs, the number of variables contributed by the receiving beamforming matrices at the relay node is $4(N-d)d$. Thus, the total number of variables is given by $8(M-d)d+4(N-d)d$.

Next, let us count the number of nulling equations. For each pair of messages, when we use an $N\times d$ receiving beamforming matrix at the relay to protect a $d$-dimensional clean subspace for that pair, we actually zero force all the signals carrying the other $6$ messages, each encoded to $d$ independent symbols. For every pair of messages, the total number of nulling equations is $6d^2$. Hence, the number of equations associated with nulling interference is given by $24d^2$.

Thus, the DoF counting bound is obtained by testing if the number of variables is no fewer than the number of nulling equations, i.e.,
\begin{eqnarray}
8(M-d)d+4(N-d)d\geq 24d^2\Longrightarrow d\leq \frac{2M+N}{9}.
\end{eqnarray}

\section{DoF Converse: Proof of Lemma \ref{lemma:x_ob}}\label{app:xob}

As implied by Lemma \ref{lemma:x_ob}, the DoF result can be represented by 4 linear pieces depending on $M$ and $N$ alternatively. For the regime $M/N\leq 2/5$, since each user equipped with $M$ antennas sends 2 messages, the DoF per message are upper bounded by $M/2$.

For the remaining regimes, we follow from the same analysis that we show for the all unicast setting in Section \ref{sec:y_ob} by deactivating the 4 messages $W_{12}=W_{21}=W_{34}=W_{43}=\emptyset$ and setting their corresponding rates $R_{12}=R_{21}=R_{34}=R_{43}=0$. Then the DoF converse proofs for $M/N\geq 5/8$, $2/5<M/N\leq1/2$ and $1/2<M/N\leq 5/8$ are essentially the same as those for $M/N\geq 7/12$, $3/7<M/N\leq1/2$ and $1/2<M/N\leq 7/12$, respectively, that we present in Section \ref{sec:y_ob}.

\section{DoF Achievability: Proof of Lemma \ref{lemma:x_ib}}\label{app:xib}

Following from the antenna dimension redundances intuition illustrated in Section \ref{sec:results}, Section \ref{sec:y_ib} and \cite{Wang_Gou_Jafar_3userMxN}, it suffice to present the achievability at $M/N=2/5$ and $M/N=5/8$. Since the DoF achievability at $M/N=5/8$, similar to that for $M/N=7/12$ of the all unicast setting, was already shown by Xiang \emph{et. al.} in \cite{Tao_sjtu}, we only study the most interesting and nontrivial case $(M,N)=(2,5)$, and show that each message has $d=1$ DoF, as implied by Theorem \ref{thm:x}. Aggain, we only need to carefully design the achievability in the first phase, so that a clean linear combination of each pair of symbols can be resolved at the relay.

In the first phase, each user $k$ encodes each symbol $u_{k\cdot}$ using a $2\times 1$ beamforming vector $V_{k\cdot}$, and the
transmitted signal vector of user $k$ can be written as
\begin{eqnarray}
X_k = V_{k3} u_{k3} + V_{k4} u_{k4},~~~~k=1,2\notag\\
X_k = V_{k1} u_{k1} + V_{k2} u_{k2},~~~~k=3,4
\end{eqnarray}
Note that there are a total of 8 symbols in this channel. Similar to our analysis for $(M,N)=(3,7)$ of the all unicast setting, in this case, at least we need to ensure that for every user, except for its 2 desired symbols and its 2 own symbols, the signal subspace occupied by the other $4$ symbols has $5-2d=3$ dimensions only. For example, consider user 1 who desires the symbols $u_{31}$ and $u_{41}$, and its own transmitted symbols are $u_{13}$ and $u_{14}$. Except for these 4 symbols, the other 4 symbols $u_{23}$, $u_{24}$ $u_{32}$ and $u_{42}$ are interfering symbols. In the $5$-dimensional space at the relay, to protect a $2$-dimensional subspace for the 2 desired symbols, those 4 vectors carrying the 4 interfering symbols sent from from user 2, user 3 and user 4 can only span $3$ dimensions. That is, we need to align one of those 4 vectors into the subspace spanned by the other 3 vectors at the relay. Among those 4 symbols, note that $u_{23}$, $u_{24}$ both originate from user 2. Since $2M<N$, every two users do not project a common intersection at the relay. Thus, we align align the vector carrying $u_{42}$ into the subspace spanned by the vectors carrying $u_{23}$, $u_{24}$ and $u_{32}$ at the relay. That is,
\begin{eqnarray}
{\bf H}_2V_{23}+{\bf H}_2V_{24}+{\bf H}_3V_{32}+{\bf H}_4V_{42}=0.\label{eqn:Xalign1}
\end{eqnarray}
Next, consider user 2 who desires the symbols $u_{32}$ and $u_{42}$, and its own transmitted symbols are $u_{23}$ and $u_{24}$. Thus, the remaining 4 symbols $u_{13}$, $u_{14}$ $u_{31}$ and $u_{41}$ are interfering symbols. In the $5$-dimensional space at the relay, in order to protect a $2$-dimensional subspace for the 2 desired symbols, we need to align one of the 4 vectors carrying those 4 symbols into the subspace spanned by the other 3 vectors at the relay. In particular, we align the vector carrying $u_{41}$ into the subspace spanned by the vectors carrying $u_{13}$, $u_{14}$ and $u_{31}$ at the relay. That is,
\begin{eqnarray}
{\bf H}_1V_{13}+{\bf H}_1V_{14}+{\bf H}_3V_{31}+{\bf H}_4V_{41}=0.\label{eqn:Xalign2}
\end{eqnarray}
With the similar analysis, for user 3, we need one alignment equation regarding the 4 interfering symbols $u_{14}$, $u_{24}$ $u_{41}$ and $u_{42}$, and we let
\begin{eqnarray}
{\bf H}_1V_{14}+{\bf H}_2V_{24}+{\bf H}_4V_{41}+{\bf H}_4V_{42}=0.\label{eqn:Xalign3}
\end{eqnarray}

Finally, for user 4, again, we need another alignment equation regarding the 4 interfering symbols $u_{13}$, $u_{23}$ $u_{31}$ and $u_{32}$. However, after carefully observing the alignment equations (\ref{eqn:Xalign1}), (\ref{eqn:Xalign2}) and (\ref{eqn:Xalign3}), we find that $(\ref{eqn:Xalign1})+(\ref{eqn:Xalign2})-(\ref{eqn:Xalign3})$ produces the following alignment equation:
\begin{eqnarray}
{\bf H}_1V_{13}+{\bf H}_2V_{23}+{\bf H}_3V_{31}+{\bf H}_3V_{32}=0.\label{eqn:Xalign4}
\end{eqnarray}
which implies that the vector carrying $u_{13}$ is automatically aligned into the subspace spanned by the vectors carrying $u_{23}$ $u_{31}$ and $u_{32}$ at the relay. That is, the equation (\ref{eqn:Xalign4}) is redundant.

Collecting the 3 equations (\ref{eqn:Xalign1}), (\ref{eqn:Xalign2}) and (\ref{eqn:Xalign3}), we rewrite them into a compact matrix form:
\begin{eqnarray}
\underbrace{\left[\!\!\begin{array}{cccccccc}{\bf O}&{\bf O}&{\bf H}_2&{\bf H}_2&{\bf O}&{\bf H}_3&{\bf O}&{\bf H}_4\\
{\bf H}_1&{\bf H}_1&{\bf O}&{\bf O}&{\bf H}_3&{\bf O}&{\bf H}_4&{\bf O}\\
{\bf O}&{\bf H}_1&{\bf O}&{\bf H}_2&{\bf O}&{\bf O}&{\bf H}_4&{\bf H}_4\end{array}\!\!\right]}_{\triangleq {\bf H}_{15\times 16}}\!\!\!
\left[\!\!\begin{array}{c}V_{13}\\ \vdots \\ V_{42}\end{array}\!\!\right]\!=\!0.\label{eqn:Xalign_c}
\end{eqnarray}
Since ${\bf H}_k,k\in\mathcal{K}$ are generic, it is not difficult to prove ${\bf H}$ has full rank, through picking a special set of matrices ${\bf H}_k,k\in\mathcal{K}$. To see this, let us pick the following special matrices:
\begin{eqnarray}
{\bf H}_1\!=\!\left[\begin{array}{cc}1&0\\0&1\\0&0\\0&0\\0&0\end{array}\right],~{\bf H}_2\!=\!\left[\begin{array}{cc}0&0\\0&0\\1&0\\0&1\\0&0\end{array}\right],~{\bf H}_3\!=\!\left[\begin{array}{cc}0&1\\1&1\\0&0\\1&0\\0&1\end{array}\right],~{\bf H}_4\!=\!\left[\begin{array}{cc}0&1\\0&0\\0&1\\0&0\\1&0\end{array}\right],\label{eqn:Xspecial}
\end{eqnarray}
and then it is easy to verify that ${\bf H}$ has full rank by computing $\det({\bf H}{\bf H}^H)\neq 0$. Also, we can obtain the beamforming vector of each symbol via solving the equation (\ref{eqn:Xalign_c}), and the solution is uniquely determined as:
\begin{eqnarray}
&&V=({\bf I}_{16}-{\bf H}^H({\bf H}{\bf H}^H)^{-1})\det({\bf H}{\bf H}^H)Q_{16\times 1},\label{eqn:XV}\\
&&\begin{array}{lll}
V_{13}=V(1:2),&V_{14}=V(3:4),\\
V_{23}=V(5:6),&V_{24}=V(7:8),\\
V_{31}=V(9:10),&V_{32}=V(11:12),\\
V_{41}=V(13:14),&V_{42}=V(15:16)
\end{array}
\end{eqnarray}
where $Q_{16\times 1}$ is a randomly picked $16\times 1$ vector. Note that the vector $V$ in (\ref{eqn:XV}), lying in the null space of ${\bf H}$, is a polynomial of all entries of ${\bf H}_k,k\in\mathcal{K}$. Thus, if we still pick the matrices in (\ref{eqn:Xspecial}), and the resulting beamforming vector of each symbol can be simplified as:
\begin{eqnarray}
\left[V_{13}~V_{14}\right]=\left[V_{23}~V_{24}\right]=\alpha\left[\begin{array}{rr}0&-1\\-1&0\end{array}\right],~~~
\left[V_{31}~V_{32}\right]=\alpha\left[\begin{array}{rr}0&1\\1&-1\end{array}\right],~~~
\left[V_{41}~V_{42}\right]=\alpha\left[\begin{array}{rr}-1&1\\0&1\end{array}\right]
\end{eqnarray}
where $\alpha$ is a non-zero scalar to meet the power constraint.

Next, we need to examine at the relay, for each pair of symbols, the other 6 symbols only occupy 4 dimensions, which are also linearly independent with the vector carrying each symbol in that pair. Take the symbols pair $(u_{13},u_{31})$ as an example, it can be seen that the vectors carrying the other 6 symbols are indeed aligned into at most 4 dimensions subspace because two alignment equations (\ref{eqn:Xalign1}) and (\ref{eqn:Xalign3}) are simultaneously satisfied. What remains to be shown is that for ${\bf H}_1V_{13}$ and ${\bf H}_3V_{31}$, each does not lie in the subspace occupied by the other 6 symbols. Since the channel matrices and beamforming vectors are already available, it is easy to verify, and we omit the calculation here.

\section{Feasibility of Linear Alignment: Proof of Theorem \ref{theorem:feasibility}} \label{app:feasibility}

In this section, we present the proof of Theorem \ref{theorem:feasibility}. For the 4-user relay MIMO $Y$ channel defined in Section \ref{sec:system}, Theorem \ref{theorem:feasibility} implies that the DoF demand per user, $d$, is feasible with linear interference alignment if and only if $d\leq \lfloor d_Y^*\rfloor$. Since the upper bound follows directly from Lemma \ref{lemma:y_ob}, we only need to provide the achievability to show $d\leq \lfloor d_Y^*\rfloor$ DoF per message are achievable using linear beamforming schemes without the need for symbol extensions in time/frequency/space. As implied by Theorem \ref{theorem:feasibility}, the feasible DoF value is again presented by four pieces, depending on either $M$ or $N$. We will consider each regime individually.

\subsection{$M/N\leq 1/2$}

The idea behind the proof for the regime $M/N\leq 1/2$ is based on reducing the number of antennas at each user and the relay to obtain a reduced network, to which we directly apply the achievable scheme designed for $M/N=3/7$.

Let us consider the regime $M/N\leq 3/7$. In this regime, our goal is to show that each message can achieve $\lfloor M/3 \rfloor$ DoF. To see this, each message is encoded to $\lfloor M/3 \rfloor$ independent symbols. At each user, we reduce the number of antennas from $M$ to $M'=3\lfloor \frac{M}{3}\rfloor$. Also, at the relay node, we reduce the number of antennas $N$ to $N'=7\lfloor \frac{M}{3} \rfloor$. This can be done since $M\geq M'$ and $N\geq N'$. By doing so, we form a new 4-user relay MIMO $Y$ channel where each user has $M'$ antennas, the relay has $N'$ antennas and $M'/N'=3/7$. Thus, $\frac{M'}{3}=\lfloor \frac{M}{3} \rfloor$ DoF can be achieved, by the use of he achievable scheme that we present in Section \ref{sec:y_ib} by replacing each one-dimensional symbol with an $\frac{M'}{3}$-dimensional symbol.

Next, consider the regime $3/7<M/N\leq 1/2$. Our goal is to show that each message can achieve $\lfloor N/7 \rfloor$ DoF. To see this, each message is encoded to $\lfloor N/7 \rfloor$ symbols. At the relay, we reduce the number of antennas from $N$ to $N'=7\lfloor \frac{N}{7}\rfloor$. Also, at each user node, we reduce the number of antennas $M$ to $M'=3\lfloor \frac{N}{7} \rfloor$. Again, this can be done due to the fact that $M\geq M'$ and $N\geq N'$. Hence, we form a new 4-user relay MIMO $Y$ channel where each user has $M'$ antennas, the relay has $N'$ antennas and $M'/N'=3/7$. Thus, $\frac{N'}{7} =\lfloor \frac{N}{7} \rfloor$ DoF can be achieved, by the use of the achievable scheme that we present in Section \ref{sec:y_ib} by replacing each one-dimensional symbol with an $\frac{N'}{7} $-dimensional symbol.

\subsection{$M/N\geq 1/2$}

The idea behind the proof for the regime $M/N\geq 1/2$ is based on reducing the number of antennas at the relay node only to obtain a reduced network, to which we directly apply the achievable scheme designed for $M/N=7/12$.

First, we consider the regime $M/N\geq 7/12$. In this regime, our goal is to show that each message can achieve $\lfloor N/6 \rfloor$ DoF. To see this, each message is encoded to $\lfloor N/6 \rfloor$ symbols. At the relay node, we reduce the number of antennas from $N$ to $N'=6\lfloor \frac{N}{6} \rfloor$. Since $M/N\geq 7/12$, we must have $M/N'\geq 7/12$ as well. Now, consider the number of common intersection projected from every two users, which is given by $2M-N'\geq 2M-N\geq \frac{N}{6}\geq \lfloor \frac{N}{6} \rfloor$, implying that we can randomly pick $\lfloor \frac{N}{6} \rfloor$ dimensions in that common intersection, along which the two signals carrying the corresponding pairwise $\lfloor \frac{N}{6} \rfloor$ symbols per user are aligned. Note that the analysis above is carried out via linear dimension counting. We still need a proof to show that is true through constructing a special specific channels. For example, assuming $m_1=\lfloor \frac{N}{6} \rfloor$, we choose the $N'\times M$ reduced channel matrices, still denoted as ${\bf H}_k,k\in\mathcal{K}$ for brevity, as ${\bf H}_k=[{\bf H}'_k~~{\bf R}_k],k\in\mathcal{K}$ where ${\bf R}_k,k\in\mathcal{K}$ are randomly generated $N'\times (M-3m_1)$ matrices and ${\bf H}'_k,k\in\mathcal{K}$ are given by
\begin{eqnarray}
{\bf H}_1'&\!\!\!\!=\!\!\!\!&\left[\begin{array}{ccc}{\bf I}_{m_1}&{\bf O}&{\bf O}\\{\bf O}&{\bf I}_{m_1}&{\bf O}\\{\bf O}&{\bf O}&{\bf I}_{m_1}\\{\bf O}&{\bf O}&{\bf O}\\{\bf O}&{\bf O}&{\bf O}\\{\bf O}&{\bf O}&{\bf O}\end{array}\right],~~~~
{\bf H}_2'=\left[\begin{array}{ccc}{\bf I}_{m_1}&{\bf O}&{\bf O}\\{\bf O}&{\bf O}&{\bf O}\\{\bf O}&{\bf O}&{\bf O}\\{\bf O}&{\bf I}_{m_1}&{\bf O}\\{\bf O}&{\bf O}&{\bf I}_{m_1}\\{\bf O}&{\bf O}&{\bf O}\end{array}\right],\notag\\
{\bf H}_3'&\!\!\!\!=\!\!\!\!&\left[\begin{array}{ccc}{\bf O}&{\bf O}&{\bf O}\\{\bf I}_{m_1}&{\bf O}&{\bf O}\\{\bf O}&{\bf O}&{\bf O}\\{\bf O}&{\bf I}_{m_1}&{\bf O}\\{\bf O}&{\bf O}&{\bf O}\\{\bf O}&{\bf O}&{\bf I}_{m_1}\end{array}\right],~~~~
{\bf H}_4'=\left[\begin{array}{ccc}{\bf O}&{\bf O}&{\bf O}\\{\bf O}&{\bf O}&{\bf O}\\{\bf I}_{m_1}&{\bf O}&{\bf O}\\{\bf O}&{\bf O}&{\bf O}\\{\bf O}&{\bf I}_{m_1}&{\bf O}\\{\bf O}&{\bf O}&{\bf I}_{m_1}\end{array}\right],
\end{eqnarray}
where each ${\bf O}$ stands for the $m_1\times m_1$ zero matrix. As a consequence, using the achievability scheme for $M/N'=7/12$, the beamforming matrix for each message at each user can be automatically determined as
\begin{eqnarray}
{\bf V}_k=\left[\begin{array}{c}{\bf I}_{3m_1}\\{\bf O}_{(M-3m_1)\times 3m_1}\end{array}\right],~~~~k\in\mathcal{K}
\end{eqnarray}
where user 1 encodes its $3m_1$ symbols for $W_{12}$, $W_{13}$ and $W_{14}$ sequentially, user 2 encodes its $3m_1$ symbols for $W_{21}$, $W_{23}$ and $W_{24}$ sequentially, user 3 encodes its $3m_1$ symbols for $W_{31}$, $W_{32}$ and $W_{34}$ sequentially, and user 4 encodes its $3m_1$ symbols for $W_{41}$, $W_{42}$ and $W_{43}$ sequentially. Therefore, the linear signal alignment solution exists almost surely.

Next, consider the regime $1/2\leq M/N\leq 7/12$. In this regime, our goal is to show that each message can achieve $\lfloor \frac{2M}{7} \rfloor$ DoF. To see this, each message is encoded to $\lfloor \frac{2M}{7} \rfloor$ symbols. At the relay node, we reduce the number of antennas from $N$ to $N'=6\lfloor \frac{2M}{7} \rfloor$. Now, consider the number of common intersection projected from every two users, which is given by $2M-N'\geq 2M-N\geq \frac{2M}{7}\geq \lfloor \frac{2M}{7} \rfloor$, implying that we have freedom to choose $\lfloor \frac{2M}{7} \rfloor$ dimensions to send $\lfloor \frac{2M}{7} \rfloor$ independent symbols per message. Again, we need a proof to show that is true through constructing a special specific channels. With the same construction of the specific channels as we present for $M/N\geq 7/12$ by letting $m_1=\lfloor \frac{2M}{7} \rfloor$, it can be seen that the linear signal alignment solution exists almost surely.


%
%
%
%
\end{document}